%% file: AA_2020_38882.tex
\documentclass[twocolumn,traditabstract]{aa}
\usepackage{graphicx,amsmath}
\usepackage{epsf}
\usepackage{txfonts}
\usepackage{longtable}
\usepackage{lscape}
\usepackage[draft]{hyperref} 
\usepackage[hyphenbreaks]{breakurl}
\usepackage{comment}
\usepackage{natbib}
\usepackage{upgreek}
\usepackage{float}
\usepackage[switch,pagewise]{lineno}
\usepackage{multirow}
\usepackage{caption}
\usepackage{lipsum}
\usepackage{threeparttable}
\usepackage{multirow}

\newcommand{\juan}[1]{{\bf #1}}
\renewcommand{\juan}[1]{#1}

\newcommand{\commentproof}[1]{{\bf \color{green}#1}}
\renewcommand{\commentproof}[1]{} 

\include{Planck}

\providecommand{\sorthelp}[1]{}

\newcommand{\kps}{km\,s$^{-1}$}

\newcommand{\vlsr}{$v_{\rm LSR}$}

\newcommand{\thorhi}{THOR-H{\sc i}}
\newcommand{\galfahi}{GALFA-H{\sc i}}

\begin{document}

\title{The history of dynamics and stellar feedback revealed by the H{\sc i} filamentary structure in the disk of the Milky Way.}
\titlerunning{H{\sc i} filamentary structure toward the Galactic plane}
    \author{
        J.\,D.~Soler$^{1}$\thanks{Corresponding author, \email{soler@mpia.de}},
        H.~Beuther$^{1}$,
        J.~Syed$^{1}$,   
        Y.~Wang$^{1}$,
        L.\,D.~Anderson$^{2}$,
        S.\,C.\,O.~Glover$^{3}$,
        P.~Hennebelle$^{4}$,
        M.~Heyer$^{5}$,
        Th.~Henning$^{1}$,
        A.\,F.~Izquierdo$^{6,7}$,
        R.\,S.~Klessen$^{2}$, 
        H.~Linz$^{1}$,
        N.\,M.~McClure-Griffiths$^{8}$,
        J.~Ott$^{9}$, 
        S.\,E.~Ragan$^{10}$,
        M.~Rugel$^{11}$,
        N.~Schneider$^{12}$,  
        R.\,J.~Smith$^{6}$,
        M.\,C.~Sormani$^{3}$,
        J.\,M.~Stil$^{13}$
        R.~Tre\ss$^{3}$, 
        J.\,S.~Urquhart$^{14}$.      
} 
\institute{
1. Max Planck Institute for Astronomy, K\"{o}nigstuhl 17, 69117, Heidelberg, Germany.\\
2. Department of Physics and Astronomy, West Virginia University, Morgantown, WV 26506, USA.\\
3. Universit\"{a}t Heidelberg, Zentrum f\"{u}r Astronomie, Institut f\"{u}r Theoretische Astrophysik, Albert-Ueberle-Str. 2, 69120, Heidelberg, Germany.\\
4. Laboratoire AIM, Paris-Saclay, CEA/IRFU/SAp - CNRS - Universit\'{e} Paris Diderot, 91191, Gif-sur-Yvette Cedex, France.\\
5. Department of Astronomy, University of Massachusetts, Amherst, MA 01003-9305, USA.\\
6. Jodrell Bank Centre for Astrophysics, School of Physics and Astronomy, University of Manchester, Oxford Road, Manchester M13 9PL, UK.\\
7. ESO, Karl Schwarzschild str. 2, 85748, Garching bei M\"{u}nchen, Germany.\\
8. Research School of Astronomy and Astrophysics, The Australian National University, Canberra, ACT, Australia.\\
9. National Radio Astronomy Observatory, PO Box O, 1003 Lopezville Road, Socorro, NM 87801, USA.\\
10. School of Physics and Astronomy, Cardiff University, Queen's Buildings, The Parade, Cardiff, CF24 3AA, UK.\\
11. Max Planck Institute for Radio Astronomy, Auf dem H\"{u}gel 69, 53121 Bonn, Germany.\\
12. Universit\"{a}t zu K\"{o}ln, I. Physikalisches Institut, Z\"{u}lpicher Str. 77, 50937, K\"{o}ln, Germany.\\
13. Department of Physics and Astronomy, University of Calgary, 2500 University Drive NW, Calgary, AB T2N 1N4, Canada.\\
14. Centre for Astrophysics and Planetary Science, University of Kent, Canterbury CT2 7NH, UK.
}
\authorrunning{Soler,~J.\,D. et al.}

\date{Received 09 JUL 2020 / Accepted 09 SEP 2020}

\abstract{
We present a study of the filamentary structure in the emission from the neutral atomic hydrogen (H{\sc i}) at 21\,cm across velocity channels in the 40\arcsec\ and 1.5-\kps\ resolution position-position-velocity cube resulting from the combination of the single-dish and interferometric observations in The H{\sc i}/OH/Recombination (THOR) line survey of the inner Milky Way.
Using the Hessian matrix method in combination with tools from circular statistics, we find that the majority of the filamentary structures in the H{\sc i} emission are aligned with the Galactic plane.
Part of this trend can be assigned to long filamentary structures that are coherent across several velocity channels.
However, we also find ranges of Galactic longitude and radial velocity where the H{\sc i} filamentary structures are preferentially oriented perpendicular to the Galactic plane.
These are located (i) around the tangent point of the Scutum spiral arm and the terminal velocities of the Molecular Ring, around $l$\,$\approx$\,28\deg\ and $v_{\rm LSR}$\,$\approx$\,100\,km\,s$^{-1}$,  (ii) toward $l$\,$\approx$\,45\deg\ and $v_{\rm LSR}$\,$\approx$\,50\,km\,s$^{-1}$, (iii) around the Riegel-Crutcher cloud, and (iv) toward the positive and negative terminal velocities.
Comparison with numerical simulations indicates that the prevalence of horizontal filamentary structures is most likely the result of the large-scale Galactic dynamics and that vertical structures identified in (i) and (ii) may arise from the combined effect of supernova (SN) feedback and strong magnetic fields.
The vertical filamentary structures in (iv) can be related to the presence of clouds from extra-planar H{\sc i} gas falling back into the Galactic plane after being expelled by SNe.
Our results indicate that a systematic characterization of the emission morphology toward the Galactic plane provides an unexplored link between the observations and the dynamical behaviour of the interstellar medium, from the effect of large-scale Galactic dynamics to the Galactic fountains driven by SNe.
}

\maketitle

\section{Introduction}\label{section:introduction}

The diffuse neutral atomic hydrogen (H{\sc i}) is the matrix within which star-forming clouds reside and the medium that takes in the energy injected by stellar winds, ionizing radiation, and supernovae \citep[see for example,][]{kulkarniANDheiles1987,dickeyANDlockman1990,kalberlaANDkerp2009,molinari2014}. 
The observation of its distribution and dynamics provides a crucial piece of evidence to understand the cycle of energy and matter in the interstellar medium \citep[ISM; for a review see][]{ferriere2001,draine2011,klessen2016}. 
In this paper, we present a study of the spatial distribution of the emission by H{\sc i} at 21\,cm using the observations with broadest dynamic range in spatial scales toward the Galactic plane available to this date.

The structure of the H{\sc i} emission in small velocity intervals has revealed a multitude of filamentary (slender or threadlike) structures, first identified in the earliest extended observations \citep[see for example,][]{weaverANDwilliams1974,heilesANDhabig1974,colomb1980}.
Many of these filaments are curved arcs that appear to be portions of small circles on the sky.
In some cases the diameters of these arc structures change with velocity in the manner expected for expanding shells \citep{heiles1979,heiles1984,McClure-Griffiths2002}.
These observations constitute clear evidence of the injection of energy into the ISM by supernova explosions \citep[see for example,][]{coxANDsmith1974,mckeeANDostriker1977,maclowANDklessen2004}.

The study of the H{\sc i} structure has been possible with the advent of single-dish surveys, such as the Galactic All-Sky Survey \citep[GASS,][]{McClure-Griffiths2009}, the Effelsberg-Bonn H{\sc i} Survey \citep[EBHIS,][]{kerp2011}, and the Galactic Arecibo L-Band Feed Array H{\sc i} survey \citep[GALFA-H{\sc i},][]{peek2018}.
Using the GALFA-H{\sc i} observations of 3,000 square degrees of sky at 4\arcmin\ resolution in combination with the Rolling Hough Transform (RHT), a technique from machine vision for detecting and parameterizing linear structures, \cite{clark2014} presented a pioneering work on the systematic analysis of H{\sc i} filamentary structures.
Using the EBHIS and GASS observations to produce a whole-sky H{\sc i} survey with a common resolution of 30\arcmin\ and applying the unsharp mask (USM), another technique from machine vision to enhance the contrast of small-scale features while suppressing large-scale ones, \cite{kalberla2016} presented a study of the filamentary structure of the local Galactic H{\sc i} in the radial velocity range $|v_{\rm LSR}|$\,$<$\,25\,\kps.
Both of these studies find a significant correlation between the elongation of \juan{the H{\sc i} filaments} and the orientation of the local interstellar magnetic field, which may be the product of magnetically-induced velocity anisotropies, collapse of material along field lines, shocks, or anisotropic density distributions
\citep[see for example,][]{lazarianANDpogosyan2000,heitsch2001a,chenANDostriker2015,inoueANDinutsuka2016,solerANDhennebelle2017,moczANDburkhart2018}.

Higher resolution H{\sc i} observations can only be achieved by using interferometric arrays.
The Galactic plane has been observed at a resolution of up to 1\arcmin\ in the Canadian Galactic Plane Survey \citep[CGPS,][]{taylor2003}, the Southern Galactic Plane Survey \citep[SGPS,][]{McClure-Griffiths2005}, and the Karl G. Jansky Very Large Array (VLA) Galactic Plane Survey \citep[VGPS,][]{stil2006}, as well as intermediate Galactic latitudes in the Dominion Radio Astrophysical Observatory (DRAO) H{\sc i} Intermediate Galactic Latitude Survey \citep[DHIGLS,][]{blagrave2017}.
Although these surveys are limited in sensitivity compared to the single-dish observations, they have been instrumental in the study of the multiphase structure of H{\sc i}, through the absorption toward continuum sources \citep[see for example,][]{strasser2007,dickey2009} and the absorption of background H{\sc i} emission by cold foreground H{\sc i} \citep[generically known as H{\sc i} self-absorption, HISA;][]{heeschen1955,gibson2000}.

Much of the H{\sc i} is observed to be either warm neutral medium (WNM) with $T$\,$\approx$\,$10^{4}$\,K or cold neutral medium (CNM) with $T$\,$\approx$\,$10^{2}$\,K \citep{heilesANDtroland2003ii}.
Detailed HISA studies of the CGPS observations reveal a population of CNM structures organized into discrete complexes that have been made visible by the velocity reversal of the Perseus arm's spiral density wave \citep{gibson2005}.
Using a combination of observations obtained with the Australia Telescope Compact Array and the Parkes Radio Telescope, \cite{McClure-Griffiths2006} reported a prominent network of dozens of hairlike CNM filaments aligned with the ambient magnetic field toward the Riegel-Crutcher cloud.
However, there has been no dedicated systematic study of the characteristics of these or other kinds of elongated structures in the H{\sc i} emission toward the Galactic plane. 

In this paper, we present a study of the elongated structures in the high-resolution observations of H{\sc i} emission in the area covered by The H{\sc i}/OH/Recombination line survey of the inner Milky Way \citep[THOR,][]{beuther2016,wang2020hi}.
We use the 40\arcsec-resolution H{\sc i} maps obtained through the combination of the single-dish observations from the Robert C. Byrd Green Bank Telescope (GBT) with the VLA D- and C-array interferometric observations made in VGPS and THOR, respectively.
We focus on the statistics of a particular property of the elongated H{\sc i} structures: its relative orientation with respect to the Galactic plane across radial velocities and Galactic longitude.

This paper is organized as follows.
We introduce the observations in Sec.~\ref{sec:observations}.
We present the method used for the characterization of the topology of the H{\sc i} emission in Sec.~\ref{sec:method} and comment on the results of our analysis in Sec.~\ref{sec:results}.
In Sec.~\ref{sec:discussion}, we discuss the observational effects, such as the mapping of spatial structure into the velocity space or ``velocity crowding'' and the HISA, in the interpretation of our results.
In Sec.~\ref{sec:mhdsims}, we explore the relationship between our observational results and the dynamical processes included in a set of numerical simulations of magnetohydrodynamic (MHD) turbulence taken from the ``From intermediate galactic scales to self-gravitating cores'' \cite[FRIGG,][]{hennebelle2018} and the CloudFactory \citep{smith2020} projects.
Finally, we present our conclusions in Sec.~\ref{sec:conclusions}.
We reserve additional analysis for a set of appendices as follows.
Appendix~\ref{app:Hessian} presents details on our implementation of the Hessian analysis, such as the selection of derivative kernels, noise masking, and spatial and velocity gridding.
Appendix~\ref{app:GALFA} provides further details on the study of the filamentary structures in the GALFA-H{\sc i} observations.
Appendix~\ref{app:othermethods} presents a comparison between our analysis method and the RHT and FilFinder methods.
Appendices~\ref{app:FRIGG} and \ref{app:CloudFactory} expand the analysis of the MHD simulations from FRIGG and CloudFactory, respectively.


\begin{figure*}[ht!]
\centerline{\includegraphics[width=0.99\textwidth,angle=0,origin=c]{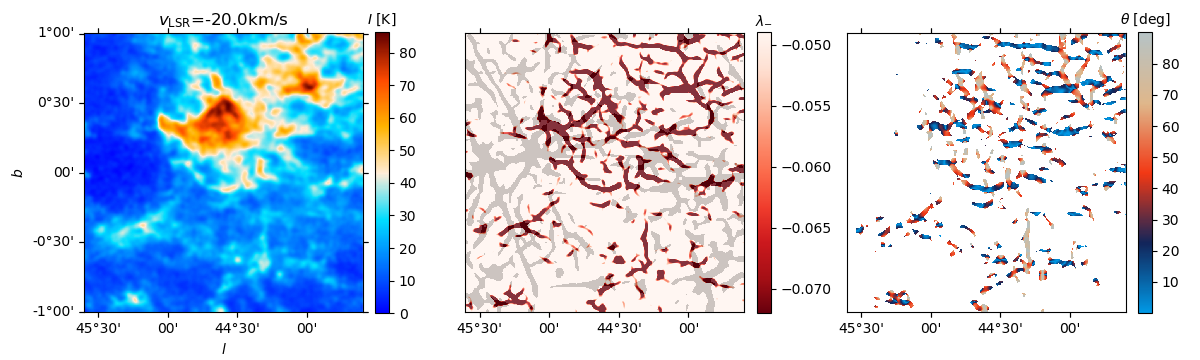}}
\vspace{-0.3cm}
\centerline{\includegraphics[width=0.99\textwidth,angle=0,origin=c]{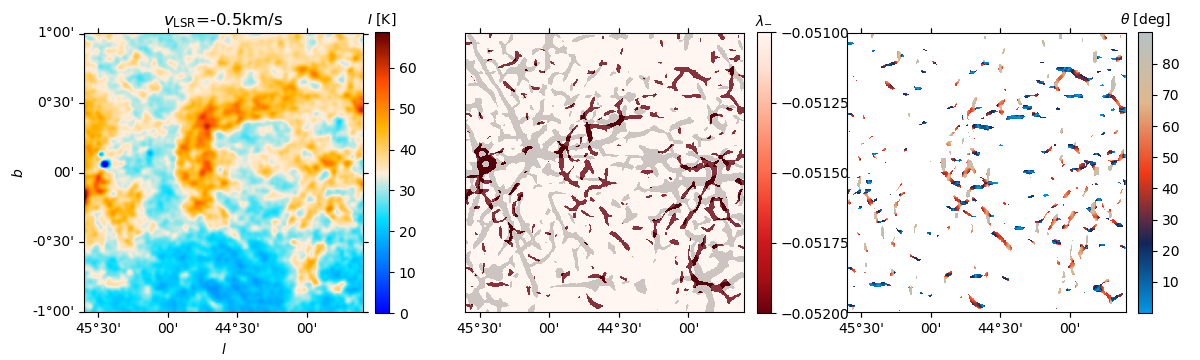}}
\vspace{-0.3cm}
\centerline{\includegraphics[width=0.99\textwidth,angle=0,origin=c]{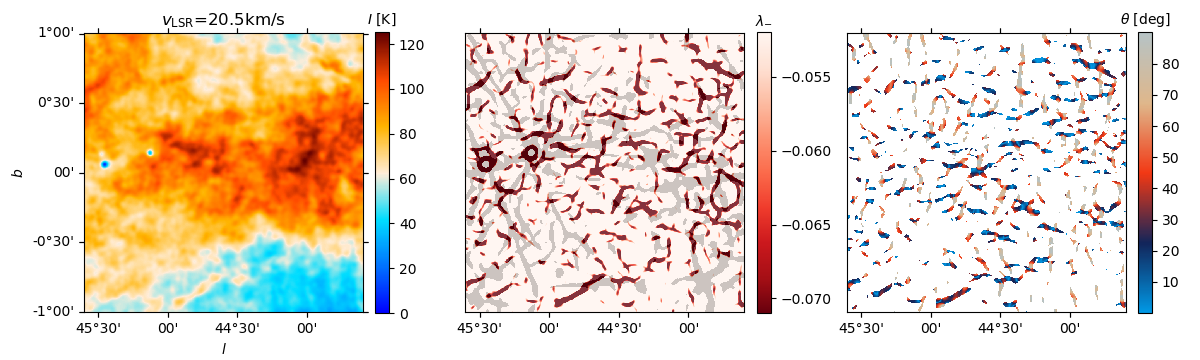}}
\caption{Examples of the Hessian analysis applied to three different velocity-channel maps toward a section of the Galactic plane observed by THOR.
{\it Left}. Intensity map.
{\it Center}. Map of the eigenvalue of the Hessian matrix identified as $\lambda_{-}$ in Eq.~\eqref{eq:lambda}, which is used to characterize the filamentary structures in the intensity map.
The overlaid grey map shows the filamentary structure obtained from the Hessian matrix analysis of the continuum noise maps, see App.~\ref{app:Hessian} for details.
{\it Right}. Map of the orientation angles evaluated using Eq.~\eqref{eq:theta}.
}\label{fig:example}
\end{figure*}

\section{Observations}\label{sec:observations}

\subsection{Atomic hydrogen emission}

For the main analysis in this paper we \juan{used} the H{\sc i} position-position-velocity (PPV) cube introduced in \cite{wang2020hi}, which we call THOR-H{\sc i} throughout this paper.
It corresponds to a combination of the single-dish observations from the GBT and the VLA C- and D-array configurations in THOR and VGPS.
The resulting data product covers the region of the sky defined by 14\pdeg0\,$\leq$\,$l$\,$\leq$\,67\pdeg0 and $|b|$\,$\leq$\,1\pdeg25 and has an angular resolution of 40\arcsec.
The THOR-H{\sc i} PPV cube, $I(l,b,v)$, \juan{was} set in Galactic coordinates and a Cartesian projection in a spatial grid with 10\arcs\,$\times$\,10\arcs\ pixels and 1.5\,\kps\ velocity channels.
\juan{It has a typical root mean square (RMS) noise of roughly 4\,K per 1.5\,\kps\ velocity channel}.
For details on the calibration and imaging of these data we refer to \cite{bihr2015,beuther2016,wang2020hi}.

We also used the GALFA-H{\sc i} observations described in \cite{peek2018} to establish a comparison with the highest resolution single-dish H{\sc i} observations.
\juan{\galfahi\ has a 4\arcm\ FHWM angular resolution.} 
\juan{At this resolution, the RMS noise of \galfahi\ and \thorhi\ per 1.5\,\kps\ velocity channel is 126 and 358\,mK, respectively.}
We re-projected the GALFA-H{\sc i} into the THOR-H{\sc i} spatial and spectral grid in two steps. 
First, we smoothed and re-gridded the data in the spectral dimension by using the tools in the {\tt spectral-cube} package \juan{({\tt http://spectral-cube.readthedocs.io})} in {\tt astropy} \citep[][]{astropy:2018}.
Second, we projected the observations into the same spatial grid of the THOR-H{\sc i} data by using the {\tt reproject} package \juan{({\tt http://reproject.readthedocs.io})}, also from {\tt astropy}.

\subsection{Carbon monoxide (CO) emission}

We compared the H{\sc i} emission observations with the $^{13}$CO\,($J$\,$=$\,1\,$\rightarrow$\,0) observations from The Boston University-Five College Radio Astronomy Observatory Galactic Ring Survey \citep[GRS,][]{jackson2006}. 
\juan{The GRS extends over most of the area covered by \thorhi, but it is limited to the} velocity range $-$5\,$\leq$\,\vlsr\,$\leq$\,135\,\kps.
\juan{It} has 46\arcsec\ angular resolution with an angular sampling of 22\arcsec\ \juan{and a velocity resolution of 0.21\,\kps.}  
\juan{Its} typical root mean square (RMS) sensitivity is 0.13\,K.
We re-projected and re-gridded the GRS data using the same procedure followed with the GALFA-H{\sc i}.
We also used the $^{12}$CO\,($J$\,$=$\,1\,$\rightarrow$\,0) presented in \citep{dame2001}.

\subsection{Catalogs of {\normalfont H}{\sc ii} regions, masers, and supernova remnants}

For the study of the relation between the orientation of the H{\sc i} filamentary structure and star formation, we used the catalogue of H{\sc ii} regions from the {\emph WISE} observations presented in \cite{anderson2014}.
Additionally, we referred to the OH masers that are also part of the THOR observations \citep{beuther2019}.
We also \juan{used} the catalogs of Galactic supernova remnants presented in \cite{anderson2017} and \cite{green2019}.

\section{Method}\label{sec:method}

Filaments, fibers, and other denominations of slender threadlike objects are common in the description and study of the ISM \citep[see for example,][]{hacar2013,andre2014,clark2019}.
In this work, we refer to the elongated structures in the H{\sc i} emission maps across velocity channels.
We characterize these structures using the Hessian matrix, a method broadly used in the study of H{\sc i} \citep{kalberla2016} and other ISM tracers \citep[see for example,][]{polychroni2013,planck2014-XXXII,schisano2014}.

The Hessian method uses the eigenvalues of the Hessian matrix at each pixel to classify them as filament-like or not.
The Hessian matrix for a given pixel is constructed by convolving the local image patch with a set of second order Gaussian derivative filters. 
Different variances of the Gaussians can be used to find filaments of various widths.
This approach does not imply that the identified structures are coherent objects in three-dimensional space, but rather aims to study the charateristics of elongated structures in the PPV space sampled by the H{\sc i} emission.

The Hessian matrix method requires a relatively low computing time, which allows for a fast evaluation of the large set of H{\sc i} observations.
It also allows for the repeated testing that is required to assess the impact of the side-lobe noise, a process that would be prohibitively time consuming using more sophisticated methods \citep[for example {\tt DisPerSE},][]{sousbie2011}.
It also yields similar results to the RHT \juan{\citep{clark2014}} and FilFinder \citep{koch2015}, as we show in App.~\ref{app:othermethods}.
Our implementation of this method is as follows.

\subsection{The Hessian matrix method}

For each position of an intensity map corresponding to $v_{\rm LSR}$\,$=$\,$v$ and a velocity interval $\Delta v$, $I(l,b,v)$\,$\equiv$\,$I(l,b,v$\,$\pm$\,$\Delta v)$, we estimate the first and second derivatives with respect to the local coordinates $(x,y)$ and build the Hessian
matrix,
\begin{equation}
\mathbf{H}(x,y) \equiv \begin{bmatrix} 
H_{xx} & H_{xy} \\
H_{yx} & H_{yy} 
\end{bmatrix},
\end{equation}
where $H_{xx}$\,$\equiv$\,$\partial^{2} I/\partial x^{2}$, $H_{xy}$\,$\equiv$\,$\partial^{2} I/\partial x \partial y$, $H_{yx}$\,$\equiv$\,$\partial^{2} I/\partial y \partial x$, and $H_{yy}$\,$\equiv$\,$\partial^{2} I/\partial y^{2}$  are the second-order partial derivatives and $x$ and $y$ refer to the Galactic coordinates $(l, b)$ as $x$\,$\equiv$\,$l\cos b$ and $y$\,$\equiv$\,$b$, so that the $x$-axis is parallel to the Galactic plane.
The partial spatial derivatives are obtained by convolving $I(l,b)$ with the second derivatives of a two-dimensional Gaussian function with standard deviation $w$.
Explicitly, we use the {\tt gaussian\_filter} function in the open-source software package {\tt SciPy}.
In the main text of this paper we present the results obtained by using a derivative kernel with 120\arcsec\ \juan{full width at half maximum (FWHM)}, which corresponds to three times the values of the beam FWHM in the THOR-H{\sc i} observations.
We also select $\Delta v$\,$=$\,1.5\,\kps\ to match the THOR-H{\sc i} spectral resolution.
The results obtained with different derivative kernel sizes and $\Delta v$ selections are presented in App.~\ref{app:Hessian}.

The two eigenvalues ($\lambda_{\pm}$) of the Hessian matrix are found by solving the characteristic equation,
\begin{equation}\label{eq:lambda}
\lambda_{\pm} = \frac{(H_{xx}+H_{yy}) \pm \sqrt{(H_{xx}-H_{yy})^{2}+4H_{xy}H_{yx}}}{2}.
\end{equation}
Both eigenvalues define the local curvature of the intensity map.
In particular, the minimum eigenvalue ($\lambda_{-}$) highlights filamentary structures or ridges, as illustrated in Fig~\ref{fig:example}.
\juan{Throughout this paper we use the word curvature to describe the topology of the scalar field $I$ and not the arc shape of some of the filaments}.

The eigenvector corresponding to $\lambda_{-}$ defines the orientation of intensity ridges with respect to the Galactic plane, which is characterized by the angle
\begin{equation}\label{eq:theta}
\theta = \frac{1}{2}\arctan\left[\frac{H_{xy}+H_{yx}}{H_{xx}-H_{yy}}\right].
\end{equation}
\juan{In practice, we estimated an angle $\theta_{ij}$ for each one of the $m$\,$\times$\,$n$ pixels in a velocity channel map, where the indexes $i$ and $j$ run over the pixels along the $x$- and $y$-axis, respectively.} 
This angle, however, is only meaningful in regions of the map that are rated as filamentary according to selection criteria based on the values of $\lambda_{-}$ and on the noise properties of the data.

\begin{figure}[ht!]
\centerline{
\includegraphics[width=0.5\textwidth,angle=0,origin=c]{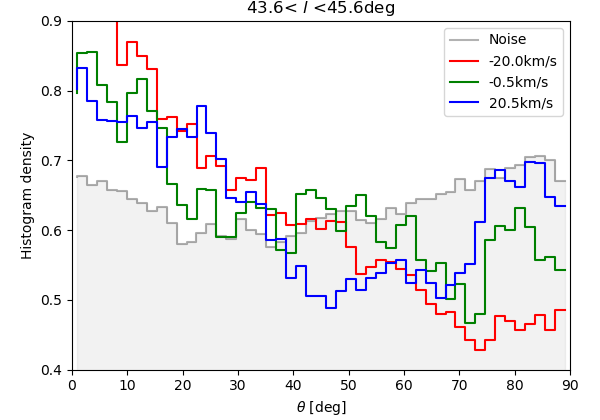}
}
\caption{Histograms of orientation angles ($\theta$, as defined in Eq.~\ref{eq:theta}) of the filamentary structures obtained by applying the Hessian matrix method to the velocity channels maps presented in Fig.~\ref{fig:example}.
The angles $\theta$\,$=$\,0\deg\ and $\theta$\,$=$\,90\deg\ correspond to the directions parallel and perpendicular to the Galactic plane, respectively. 
}\label{fig:exampleHistoTheta}
\end{figure}

\begin{figure}[ht!]
\centerline{
\includegraphics[width=0.5\textwidth,angle=0,origin=c]{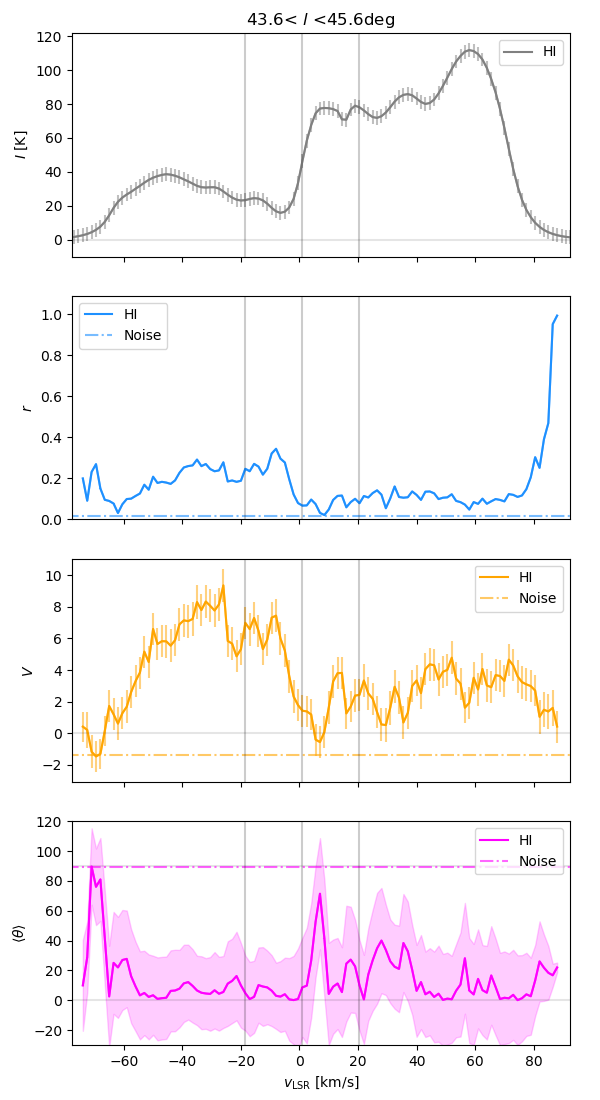}
}
\caption{Mean spectrum and circular statistics used for the study of the orientation of the H{\sc i} filamentary structures across velocity channels toward the region presented in Fig.~\ref{fig:example}. 
{\it Top}. Mean H{\sc i} intensity ($I$).
{\it Middle top}. Mean resulting vector ($r$), which indicates if the distribution of orientation angles is flat ($r$\,$\approx$\,0) or sharply unimodal ($r$\,$\approx$\,1).
{\it Middle bottom}. Projected Rayleigh statistic ($V$), which indicates if the distribution of orientation angles is clearly peaked around 0\deg\ ($V$\,$\gg$\,\juan{0}) or 90\deg\ ($V$\,$\ll$\,\juan{0}).
{\it Bottom}. \juan{Circular mean ($\left<\theta\right>$) and standard deviation ($\sigma_{V}$) of the orientation angles, represented by the line and the shaded area, respectively.}
\juan{The error bars correspond to the observation uncertainties and the error propagation based on the Monte Carlo sampling introduced in App.~\ref{app:MCs}.}
The dashed horizontal lines in each panel indicate the values of those quantities in the 1.4\,GHz noise maps used to characterize the effect of the side lobes and continuum sources in the H{\sc i} maps.
The vertical gray lines correspond to the velocity channels shown in Fig.~\ref{fig:example}.
}\label{fig:exampleCircStats}
\end{figure}
\begin{figure}[ht!]
\centerline{
\includegraphics[width=0.5\textwidth,angle=0,origin=c]{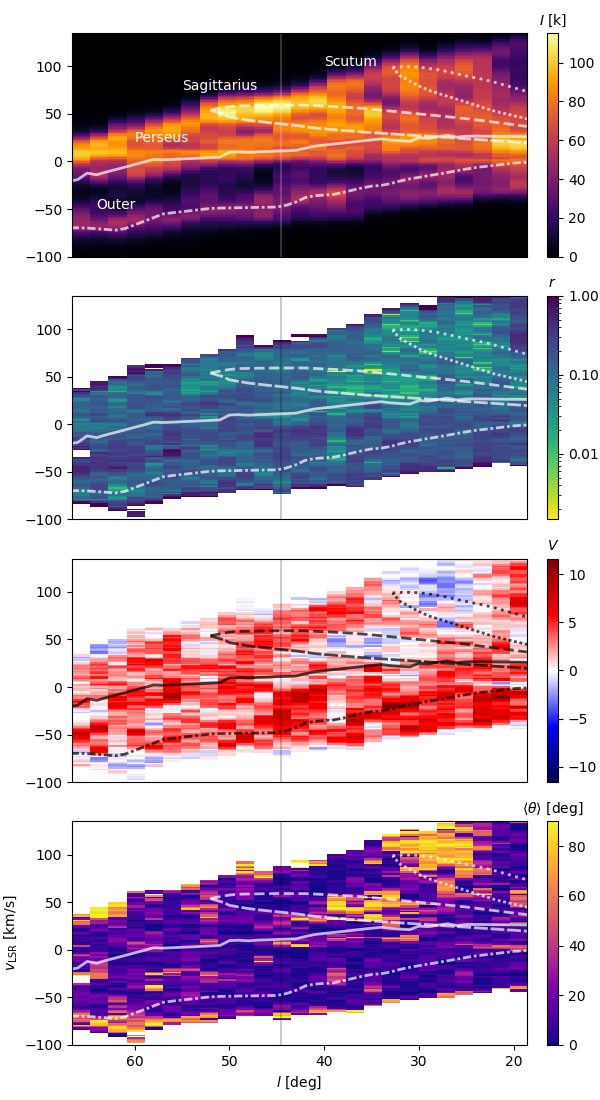}
}
\caption{Longitude-velocity ($lv$) diagrams corresponding to the mean intensity and the results of the Hessian analysis of the velocity-channel maps in the region covered by the THOR observations. 
Each pixel in the diagrams represents the quantities calculated in 2\deg\,$\times$\,2\deg\ tile in a 1.5\,\kps-velocity channel as follows. 
{\it Top}. Mean H{\sc i} intensity ($I$).
{\it Middle top}. Mean resulting vector ($r$).
{\it Middle bottom}. Projected Rayleigh statistic ($V$).
{\it Bottom}. Mean orientation angle ($\left<\theta\right>$) of the H{\sc i} filamentary structures, which is only well-defined if $r$\,$>$\,0.
The overlaid curves correspond to selected spiral arms from the model presented in \cite{reid2016}.
The vertical line indicates the central $l$ of the 2\deg\,$\times$\,2\deg\ tile presented in Fig.~\ref{fig:example} and whose circular statistics are shown in Fig.~\ref{fig:exampleCircStats}.
}\label{fig:lvdiagrams}
\end{figure}

\subsection{Selection of the filamentary structures}\label{sec:filselection}

We applied the Hessian matrix analysis to each velocity channel map, as illustrated in Fig.~\ref{fig:example}, but then selected the regions that are significant for the analysis by following three criteria.
The first selection of filamentary structures is based on the noise properties of the H{\sc i} observations.
We selected regions where $I(l,b)$\,$>$\,5$\sigma_{I}$, where $\sigma_{I}$ is approximately 4\,K and is estimated from the standard deviation of $I$ in \juan{noise-dominated} velocity channels. 

The second selection criterion addresses the fact that the noise in the interferometric data is affected by the artifacts resulting from residual side lobes with amplitudes that vary depending on the sky position.
The side lobes can introduce spurious filamentary structures in the H{\sc i} emission around continuum sources, which are seen in absorption.
To mitigate this effect, we mask regions of the map by using a threshold on the continuum emission noise maps introduced in \cite{wang2018}, as detailed in App.~\ref{app:Hessian}.
For the sake of illustration, we include the orientation of the noise features characterized with the Hessian method in the examples presented in Figs.~\ref{fig:exampleHistoTheta} and \ref{fig:exampleCircStats}, which correspond to a 2\deg\,$\times$\,2\deg\ tile toward the center of the observed Galactic longitude range.

The final selection criterion is based on the values of the eigenvalue $\lambda_{-}$.
Following the method introduced in \cite{planck2014-XXXII}, we estimate this quantity in velocity channels dominated by noise.
For that purpose, we select the five \juan{noise-dominated} velocity channels \juan{and estimate the minimum value of $\lambda_{-}$ in each one of them}.
\juan{We use the median of these five $\lambda_{-}$ values as the threshold value, $\lambda^{C}_{-}$}.
\juan{The median aims to reduce the effect of outliers, but in general the values of $\lambda_{-}$ in the noise-dominated channels are similar within a 10\% level and this selection does not imply any loss of generality.}
We exclusively consider regions of each velocity channel map where $\lambda_{-}$\,$<$\,$\lambda^{C}_{-}$, given that the most filamentary structure has more negative values of $\lambda_{-}$. 

\subsection{Characterization of the filament orientation}\label{sec:circstats}

Once the filamentary structures are selected, we use the angles estimated using Eq.~\ref{eq:theta} to study their orientation with respect to the Galactic plane, as illustrated in Fig.~\ref{fig:example}.
The histograms presented in Fig.~\ref{fig:exampleHistoTheta} show the variation of the preferential orientation across velocity channels.
For a systematic evaluation of this variation, we use three quantities commonly used in the field of circular statistics \citep[see for example,][]{batschelet1981}: the mean resultant vector ($r$), the projected Rayleigh statistic ($V$), and the mean orientation angle $\left<\theta\right>$\juan{, which are calculated using the functions in the {\tt magnetar} package} (\burl{https://github.com/solerjuan/magnetar}).

These statistical tools provide different and complementary information on the distribution of orientation angles, for example, $r$ indicates if the distribution is flat or there is a preferred orientation.
The values of $V$ indicate the significance of that preferred orientation with respect to two directions of interest, namely, parallel or perpendicular to the Galactic plane, that is, 0 and 90\deg.
Finally, $\left<\theta\right>$ indicates if the preferred orientation is different to the reference directions indicated in the computation of $V$, although this can be an ill-defined quantity if the angle distribution is random or multimodal ($r$\,$\approx$\,0).
\juan{Figure~\ref{fig:exampleCircStats} presents an example of the values of these three quantities across velocity channels toward the tile presented in Fig.~\ref{fig:example}.}

\subsubsection{Is there a preferred orientation of the H{\sc i} filaments?}

The mean resultant vector is defined as 
\begin{equation}\label{eq:mymrv}
r=\frac{\left(\left[\sum^{m,n}_{ij}w_{ij}\cos(2\theta_{ij})\right]^{2}+\left[\sum_{ij}w_{ij}\sin(2\theta_{ij})\right]^{2}\right)^{1/2}}{\sum_{ij}w_{ij}},
\end{equation}
where the indices $i$ and $j$ run over the pixel locations in the two spatial dimensions $(l,b)$ for a given velocity channel and $w_{ij}$ is the statistical weight of each angle $\theta_{ij}$.
We account for the spatial correlations introduced by the telescope beam by choosing $w_{ij}$\,$=$\,$(\delta x/\Delta)^{2}$, where $\delta x$ is the pixel size and $\Delta$ is the diameter of the derivative kernel that we use to calculate the gradients, if it is larger than the beam size.
If it is smaller than the beam size, the main scale of spatial correlation is that of the beam and consequently $\Delta$ should correspond to the beam size.
We note that the statistical weights cancel out in Eq.~\eqref{eq:mymrv}, but we include them for the sake of completeness and because they are crucial to evaluate the significance of $V$ \citep[see for example,][]{fissel2019,soler2019hro,heyer2020}.

The mean resultant vector, $r$, is a descriptive quantity of the angle distribution that can be interpreted as the percentage of filaments pointing in a preferential direction.
If $r$\,$\approx$\,0, the distribution of angles is either uniform or on average does not have a well-defined mean orientation.
Consequently, $r$\,$\approx$\,1 only if almost all of the angles are very similar.

\juan{The uncertainty on $r$ is related to the uncertainty on the orientation angles estimated with Eq.~\eqref{eq:theta}.
Given the selection of the filamentary structures based on the $I$ signal-to-noise ratio, presented in Sec.~\ref{sec:filselection}, and the additional spatial smoothing introduced by the derivative kernel, the uncertainty on $r$ is expected to be very small.
This assumption is confirmed by the results of the Monte Carlo sampling introduced in App.~\ref{app:MCs}, which leads to the imperceptible error bars in Fig.~\ref{fig:exampleCircStats}.}

\subsubsection{What is the significance of the orientations parallel or perpendicular to the Galactic plane?}

We also considered the projected Rayleigh statistic ($V$), a test to determine whether the distribution of angles is non-uniform and peaked at \juan{a particular angle}.
\juan{This test is equivalent to the modified Rayleigh test for uniformity proposed by \cite{durandANDgreenwood1958}, for the specific directions of interest $\theta$\,$=$\,$0$\deg\ and 90\deg\ \citep{jow2018}}.
It is defined as
\begin{equation}\label{eq:myprs}
V = \frac{\sum^{n,m}_{ij}w_{ij}\cos(2\theta_{ij})}{\sqrt{\sum^{n,m}_{ij}w_{ij}/2}},
\end{equation}
which follows the same conventions introduced in Eq.~\eqref{eq:mymrv}.

\juan{In the particular case of independent and uniformly distributed angles, and for large number of samples, values of $V$\,$=$\,$1.64485$ and $2.5752$ correspond to statistical significance levels of 5\% and 0.5\%, respectively \citep{durandANDgreenwood1958,batschelet1972}.
Values of $V$\,$\approx$\,1.71 and 2.87 are roughly equivalent to 2- and 3-$\sigma$ confidence intervals.
We note, however, that the correlation across scales in the ISM makes it very difficult to estimate the absolute statistical significance of $V$.
In our application, we accounted for the statistical correlation brought in by the beam by introducing the statistical weights $w_{ij}$.
Throughout this paper we considered values of $V$\,$>$\,$3$ or $V$\,$<$\,$-3$ as a compelling indication of a preferred orientation of 0\deg\ or 90\deg, respectively}.

\juan{The uncertainty on $V$ can be estimated by assuming that each orientation angle $\theta_{ij}$ derived from Eq.~\eqref{eq:theta} is independent and uniformly distributed, which leads to the bounded function
\begin{equation}\label{eq:s_prs}
\sigma^{2}_V = \frac{2\sum^{n}_{ij}w_{ij}\left[\cos(2\theta_{ij})\right]^{2}-V^{2}}{\sum_{ij}w_{ij}},
\end{equation}
as described in \cite{jow2018}.
In the particular case of identical statistical weights $w_{ij}$\,$=$\,$w$, $\sigma^{2}_V$ has a maximum value of $w$.
We also estimated $\sigma^{2}_V$ by directly propagating the Monte Carlo sampling introduced in App.~\ref{app:MCs}.
This method produces slightly higher values than those found using Eq.~\eqref{eq:s_prs}, most likely because it accounts for the correlation between the orientation angles in the map.
Thus, we report the results of the latter method in the error bars in Fig.~\ref{fig:exampleCircStats}.}

\subsubsection{Is there another mean orientation different to parallel or perpendicular to the Galactic plane?}

For the sake of completeness, we also computed the mean orientation angle, defined as
\begin{equation}\label{eq:meanangle}
\left<\theta\right>\,\equiv\,0.5\arctan\left[\frac{\sum_{ij}w_{ij}\sin(2\theta_{ij})}{\sum_{ij}w_{ij}\cos(2\theta_{ij})}\right].
\end{equation}
This quantity highlights orientations that are not considered in the statistical test $V$, that is, \juan{different to} 0\deg\ or 90\deg.

\juan{As in the case of $r$, the selection of the structures introduced in Sec.~\ref{sec:filselection} guarantees very small uncertainties on the values $\left<\theta\right>$.
There is, however, important information in the standard deviation of the angle distribution, which corresponds to the circular variance
\begin{equation}
\sigma_{V}\equiv\sqrt{-2\ln(r)}
\end{equation}
\citep[see for example][]{batschelet1981}, which we also report in Fig.~\ref{fig:exampleCircStats}.
}

\juan{We found that the} values of $r$, $V$, and $\left<\theta\right>$ change across velocity channels, {as we show in Fig.~\ref{fig:exampleCircStats} for the example region presented in Fig.~\ref{fig:example}.}
\juan{The} values of $r$ are on average around 0.1, which means that the signal of a preferential orientation in the H{\sc i} filamentary structure corresponds to roughly 10\% of the selected pixels in each velocity channel map.
A clear exception are the channels close to the maximum and minimum $v_{\rm LSR}$, where the values of $r$ are large, a result that we \juan{considered in more detail} in Sec.~\ref{sec:results}.

\juan{The values of $V$\,$>$\,$0$ in Fig.~\ref{fig:exampleCircStats} imply that the H{\sc i} filamentary structures are mostly parallel to the Galactic plane.
This result is further confirmed by the values of $\left<\theta\right>$, which are close to 0\deg.
We found that roughly half of the velocity channels have values of $V$\,$>$\,$3$, which corresponds to a 3-$\sigma$ confidence interval.}
\juan{The trends in $V$ and $\left<\theta\right>$ from the H{\sc i} emission} are significantly different from that of the \juan{continuum} noise, \juan{which we used as a proxy of the artifacts introduced by strong continuum sources}.
\juan{They} are \juan{also} independent of the average intensity in each velocity channel\juan{, which implies that} they are a result of the \juan{spatial distribution of the emission and not the amount of it.}

\section{Results}\label{sec:results}

\subsection{Longitude-velocity diagrams}

We applied the Hessian matrix analysis \juan{with a derivative kernel of 120\arcs\ FWHM} to the whole area covered by the THOR-H{\sc i} observations by evaluating 2\deg\,$\times$\,2\deg\ non-overlapping tiles for which we estimated $r$, $V$, and $\theta$ across velocity channels.
The 2\deg\,$\times$\,2\deg\ area provides enough spatial range for the evaluation of the filamentary structures.
This selection does not significantly affect our results, as further described in App.~\ref{app:Hessian}.

The results of the Hessian analysis of the tiles are summarized in the longitude-velocity ($lv$) diagrams presented in Fig.~\ref{fig:lvdiagrams}.
The empty (white) portions of the $lv$ diagrams correspond to tiles and velocity channels \juan{discarded by} the selection criteria introduced in Sec.~\ref{sec:filselection}.
\juan{The color map for $V$ was chosen such that the white color also corresponds to $V$\,$\approx$\,0, where the results of this statistical test are inconclusive.}

We note that the values of $r$ around the maximum and minimum velocities are particularly high with respect to most of the other tiles.
The fact that these high values of $r$ appear in velocity channels where $\left<I\right>$ is low indicates that the circular statistics are dominated by just a few filamentary structures that are above the $I$ threshold.
For example, if there is just one straight filament in a tile, $r$\,$=$\,1.
If there are ten straight filaments in a tile, all of them would have to be parallel to produce $r$\,$=$\,1.
Thus, the high values of $r$ are a product of the normalization of this quantity rather than due to a lack of significance in other areas.
Despite this feature, $r$ is a powerful statistic that indicates that, on average across $l$ and $v_{\rm LSR}$, the preferential orientations correspond to around 17\% of the H{\sc i} filamentary structures. 

The values of $|V|$\,$\gg$\,\juan{0} in the middle panel of Fig.~\ref{fig:lvdiagrams}, indicate that the H{\sc i} filaments are preferentially orientated either parallel or perpendicular to the Galactic plane.
This is further confirmed by the values of $\left<\theta\right>$, shown in the bottom panel of Fig.~\ref{fig:lvdiagrams}.
However, it does not imply that other orientations different from 0 and 90\deg\ are not present in some tiles and velocity channels or that the distribution of the orientation of H{\sc i} filaments is bimodal, as detailed in App~\ref{app:GALFA}.
Figure~\ref{fig:lvdiagrams} shows that the most significant trend is for the H{\sc i} filaments to be parallel to the Galactic plane. 
\juan{T}he 90\deg\ orientation is significant because it is clearly grouped in specific ranges of $l$ and $v_{\rm LSR}$.

\begin{figure*}[ht!]
\centerline{\includegraphics[width=0.99\textwidth,angle=0,origin=c]{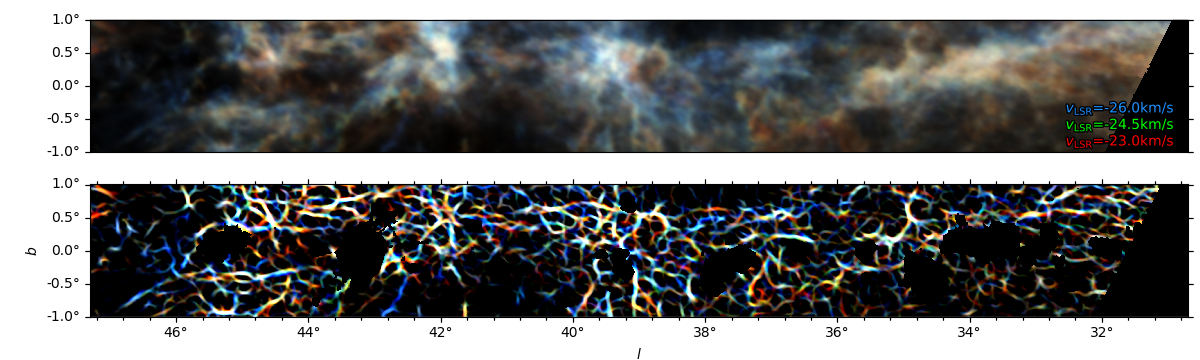}}
\centerline{\includegraphics[width=0.99\textwidth,angle=0,origin=c]{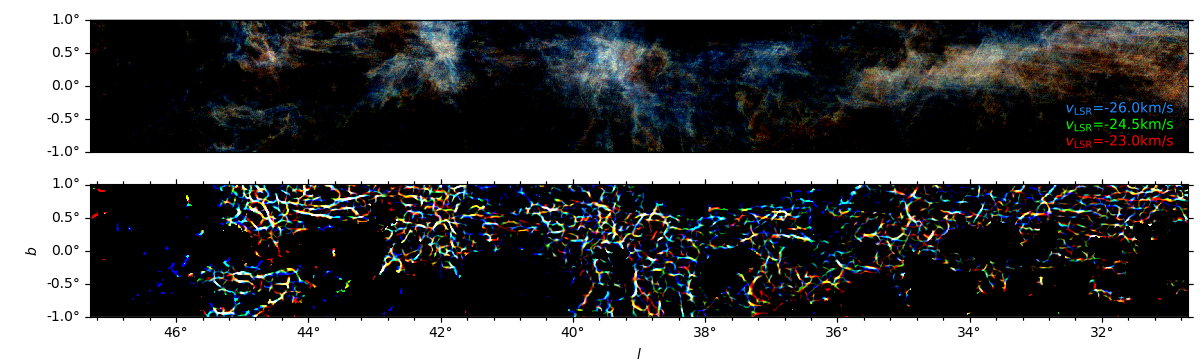}}
\caption{GALFA-H{\sc i} (top) and THOR-H{\sc i} (bottom) observations and their respective filaments from the Hessian analysis toward the same region of the Galactic plane in the indicated velocity channels.}
\label{fig:THORandGALFAcomparisonRGB}
\end{figure*}

\begin{figure}[ht!]
\centerline{
\includegraphics[width=0.49\textwidth,angle=0,origin=c]{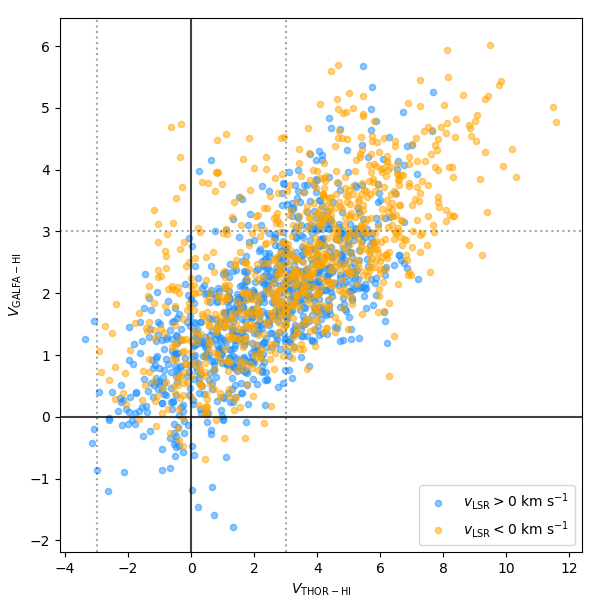}
}
\caption{Projected Rayleigh statistic $V$ calculated from Hessian analysis of the \thorhi\ and \galfahi\ observations using a 120\arcsec\ FWHM derivative kernel.
Each point corresponds to a \juan{2\deg\,$\times$\,2\deg\,$\times$\,1.5\,\kps\ velocity-channel tile fully covered by both surveys}.
\juan{The vertical and horizontal dotted lines indicate $|V|$\,$=$\,3, which roughly corresponds to a 3-$\sigma$ confidence level in the estimation of a preferential orientation of 0\deg\ or 90\deg.}
The difference in the range of $V$ is the result of the larger significance obtained with the higher-resolution observations.
}
\label{fig:scatterTHORandGALFA_PRS}
\end{figure}

\subsection{Comparison to previous observations}

We compare\juan{d} the results of our study with the analysis of the \galfahi\ observations, which do not cover the entire area of the THOR survey but provide a benchmark dataset to test the possible artifacts introduced by the side lobe noise in the interferometric observations.
Figure~\ref{fig:THORandGALFAcomparisonRGB} presents an example of the Hessian analysis applied to the same region and velocity channels in \galfahi\ and \thorhi.
Visual inspection of these and other regions confirms that the structures selected in \thorhi\ have a counterpart in \galfahi.
In general, structures that appear monolithic in GALFA-H{\sc i} are decomposed into smaller components, and filamentary structures appear to be narrower in \thorhi.
Moreover, the general orientation of the structures is very similar, further confirming that the filamentary structures in THOR-H{\sc i} are not produced by the side lobe noise.

The direct comparison of the $V$ values in both datasets, shown in Fig.~\ref{fig:scatterTHORandGALFA_PRS}, provides a quantitative evaluation of the differences in the orientation of the filaments in GALFA-H{\sc i} and THOR-H{\sc i}. 
It is evident that there is a linear correlation between the values of $V$ in both datasets, as expected from the common scales in the observations.
This trend is offset from the direct one-to-one relation, \juan{for example, the tiles that present $V$\,$\approx$\,2 in GALFA-H{\sc i} show $V$\,$\approx$\,4 in THOR-H{\sc i}}.
\juan{This increase in the values of $V$ is related to the higher number of independent gradients that is obtained with the higher resolution THOR-H{\sc i} observations, as detailed in App.~\ref{app:GALFA}.
The factor of two is consistent with the square root of the ratio between the statistical weights used for each dataset, which following Eq.~\eqref{eq:myprs} is equivalent to the ratio between the spatial scales at which we evaluated the filamentary structures, namely, the kernel size of 120\arcs, and the GALFA-H{\sc i} 240\arcs\ beam}.

The positive values of $V$ in Fig.~\ref{fig:scatterTHORandGALFA_PRS} indicate that the general trend across $l$ and $v_{\rm LSR}$ is roughly the same in both datasets, that is, filamentary structures in the H{\sc i} emission run mostly parallel to the Galactic plane.
Unfortunately, the region of the $lv$-diagram with the most prominent grouping of $V$\,$<$\,0 in THOR-H{\sc i}, $l$\,$\approx$\,27\deg\ and $v_{\rm LSR}$\,$\approx$\,100\,\kps, is not covered by the GALFA-H{\sc i} observations.
However, the general agreement obtained in the region covered by both surveys indicates that the global results of the Hessian analysis of THOR-H{\sc i} are not the product of potential artifacts in the interferometric image reconstruction.
We provide further details on the comparison between the \galfahi\ and \thorhi\ in App.~\ref{app:GALFA}

\subsection{General trends}

The general trend revealed by the Hessian analysis of the H{\sc i} emission across velocity channels is the presence of filamentary structures that tend to be preferentially parallel to the Galactic plane, as illustrated in Fig.~\ref{fig:lvdiagrams}.
Below, we detail some observational considerations that are worth highlighting.

First, there is a significant preferential orientation of the filamentary structures in the H{\sc i}.
This holds despite the fact that many of the density structures in position-position-position (PPP) space are crammed into the same velocity channel in PPV, an effect called velocity crowding \citep[see for example,][]{beaumont2013}.
Figure~\ref{fig:scatterTHORandGALFA_PRS} shows \juan{similar orientation trends in the ranges \vlsr\,$<$\,0\,\kps\ and \vlsr\,$>$\,0\,\kps, although there is a larger spread in $V$ in the range \vlsr\,$<$\,0\,\kps}.
\juan{This is presumably related to the velocity crowding due to the mapping of at least two positions into the same radial velocity, which is more pronounced at \vlsr\,$>$\,0\,\kps}, under the assumption of circular motions \citep{reid2014}.
\juan{The velocity crowding is most likely averaging the orientation of the H{\sc i} filaments and eliminating the outliers seen at \vlsr\,$<$\,0\,\kps}.
\juan{This averaging is not producing a random distribution of orientation but rather a tighter distribution around 0\deg}.
If velocity crowding had completely randomized the orientation of the H{\sc i} filaments, we would see a clearer tendency toward $V$\,$\approx$\,0 in the \vlsr\,$>$\,0\,\kps\ velocity range{, which is clearly not the case in Fig.~\ref{fig:scatterTHORandGALFA_PRS}}.

Given that we are using square tiles in $l$ and $b$ for our analysis, it is unlikely that the prevalence of horizontal structures is biased by the aspect ratio of the analysis area, although several filaments can extend from one tile to another.
Additionally, given that we are using the same curvature criterion for all velocity channels in a particular $l$-range, as discussed in Sec.~\ref{sec:filselection}, we can guarantee that the filamentary structures are selected independently of the intensity background in a particular velocity channel.

Second, the preferential orientation of the H{\sc i} filamentary structures is persistent across \juan{contiguous} velocity channels, as illustrated in the example shown in Fig.~\ref{fig:exampleCircStats}.
This implies that there is a coherence in the structure of these filamentary structures in PPV space and they are not dominated by fluctuations of the velocity field on the scale of a few kilometers per second.
This is a significant fact given the statistical nature of our study, which characterizes the curvature of the emission rather than looking for coherent structures in PPV.

Third, there are exceptions to the general trend and they correspond to H{\sc i} filaments that are perpendicular to the Galactic plane.
These are grouped in Galactic longitude and velocity around $l$\,$\approx$\,27\deg\ to 30\deg\ and $v_{\rm LSR}$\,$\approx$\,100\,\kps\ and $l$\,$\approx$\,36\deg\ and $v_{\rm LSR}$\,$\approx$\,40\,\kps.
The fact that these velocity and Galactic longitude ranges correspond to those of dynamic features in the Galactic plane and active sites of star formation is further discussed in Sec.~\ref{sec:discussion}.

Fourth, there is a deviation of the relative orientation of the H{\sc i} filaments from the general trend, $V$\,$>$\,0, in the channels corresponding to the terminal velocities, $v_{t}$.
Inside the solar circle, each line of sight has a location where it is tangent to a circular orbit about the Galactic center. 
At this location, known as the tangent point, the projection of Galactic rotational velocity, onto the local standard of rest (LSR) is the greatest, and the measured $v_{\rm LSR}$ is called the terminal velocity.
Assuming that azimuthal streaming motions are not too large, the maximum velocity of H{\sc i} emission can be equated to the terminal velocity, $v_{t}$ \citep[see for example,][]{McClure-Griffiths2016}.

The velocity range around $v_{t}$ is the most affected by the effect of velocity crowding, that is, a velocity channel in the PPV space corresponds to multiple positions in the PPP space.
Thus, velocity crowding is a plausible explanation for the $V$ and $\left<\theta\right>$ values found toward the maximum and minimum velocities in Fig.~\ref{fig:lvdiagrams}, which deviates from the general trend $V$\,$\gg$\,0 and $\left<\theta\right>$\,$\approx$\,0\deg.
However, velocity crowding does not provide a conclusive explanation for the prevalence of vertical H{\sc i} filaments, $V$\,$\ll$\,0 and $\left<\theta\right>$\,$\approx$\,90\deg, found around the maximum and minimum velocities toward 55\pdeg0\,$\lesssim$\,$l$\,$\lesssim$\,65\pdeg0.

\section{Discussion}\label{sec:discussion}

\begin{figure*}[ht!]
\centerline{\includegraphics[width=0.99\textwidth,angle=0,origin=c]{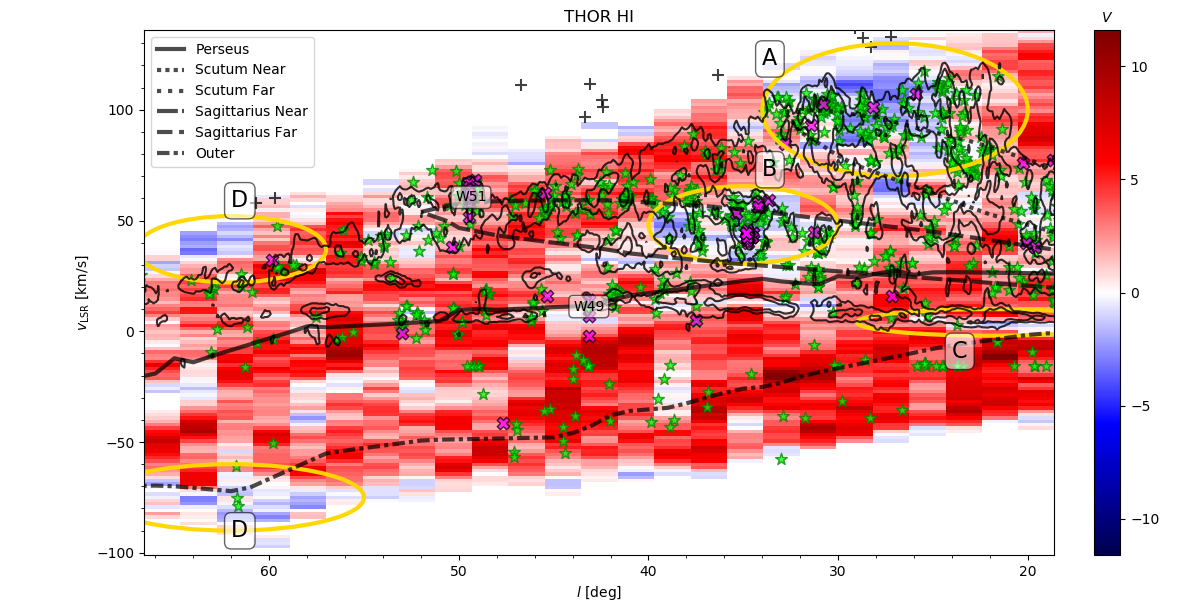}}
\caption{Projected Rayleigh statistic ($V$) corresponding to the orientation of the filamentary structures identified with the Hessian matrix method in the H{\sc i} emission in 2\deg\,$\times$\,2\deg\ tiles and across velocity channels.
The overlaid curves correspond to selected spiral arms from the model presented in \cite{reid2016}.
The black contours corresponds to the 97 and 99 percentile of the $^{12}$CO emission in the observations presented in \cite{dame2001} integrated over the Galactic longitude range $|b|$\,$\leq$\,1\deg.
The green stars correspond to the H{\sc ii} regions in the WISE catalog of Galactic H{\sc ii} regions \citep{anderson2014}.
The two most prominent regions of high-mass star formation in this portion of the Galaxy, W49 and W51, are indicated by the corresponding labels.
The magenta crosses correspond to the 1720-MHz OH masers in the catalog \citep{beuther2019}, which are typically excited by shocks.
The black crosses correspond to the compact H{\sc i} clouds at high forbidden velocities identified in the VGPS \citep{stil2006b}.
The ellipses labelled with letters indicate the regions of interest (ROIs) discussed in Sec.~\ref{sec:ROIs}.
}
\label{fig:PRSandHIIregions}
\end{figure*}

We articulate the discussion of the physical phenomena related to the results presented in Sec.~\ref{sec:results} as follows.
First, we discuss the general trend in the orientation of H{\sc i} filamentary structures. which is being parallel to the Galactic plane.
Then, we focus on the areas of the $lv$-diagram dominated by H{\sc i} structures perpendicular to the Galactic plane, identified as regions of interest (ROIs) in the $lv$-diagram presented in  Fig.~\ref{fig:PRSandHIIregions}.

\subsection{Filamentary structures parallel to the Galactic plane}

\begin{figure}[ht!]
\centerline{\includegraphics[width=0.49\textwidth,angle=0,origin=c]{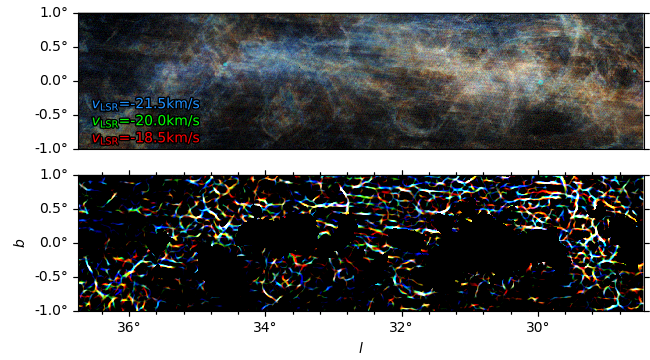}}
\caption{THOR-H{\sc i} observations (top) and their respective filaments from the Hessian analysis (bottom) toward $l$ and $v_{\rm LSR}$ ranges dominated by filamentary structures parallel to the Galactic plane.
}\label{fig:ROIparallel}
\end{figure}

Most of the filamentary structures identified with the Hessian method in the H{\sc i} emission across velocity channels are parallel to the Galactic plane.
This is hardly a surprise considering that this is the most prominent axis for an edge-on disk, but it confirms that in the observed area other effects that break this anisotropy, such as the expansion of SN remnants, do not produce significant deviations from this preferential direction.
Most of the deviations from this general \juan{trend} appear to be concentrated in the $l$ and $v_{\rm LSR}$ ranges that we discuss in the next sections.
 
However, not all of the H{\sc i} filamentary structures that are parallel to the Galactic plane appear to have the same morphology.
Figure~\ref{fig:ROIparallel} shows an example of three velocity channels with high values of $V$, which correspond to being significantly dominated by H{\sc i} filaments parallel to the Galactic plane.
We note a few narrow structures that appear to be randomly oriented, which can be related to the fibers identified in \cite{clark2014}, and even some prominent vertical filaments, but the prevalent structures are parallel to the Galactic plane.
These extend over several degrees in $l$ and appear to have a width of at least 0\pdeg5, although they are decomposed in smaller segments that correspond to the 120\arcs\ width of the derivative kernel (see App.~\ref{app:Hessian} for a discussion on the selection of the kernel size).

The predominantly positive values of $V$ observed at \vlsr\,$<$\,0\,\kps\ can be associated with a distance effect.
Assuming circular motions, \vlsr\,$<$\,0\,\kps\ roughly corresponds to distances larger than 10\,kpc, where the apparent size of the H{\sc i} disk fits in the THOR-H{\sc i} $b$ range. 
This would appear as a concentration of brightness that could bias $r$ to higher values and $V$ to more positive values, as illustrated in Fig.~\ref{fig:exampleCircStats}.
Yet we observe vertical H{\sc i} filaments at \vlsr\,$<$\,0\,\kps, as shown in Fig.~\ref{fig:THORandGALFAcomparisonRGB}.

At shorter distances, the horizontal structures in H{\sc i} emission are reminiscent of the filamentary structures parallel to the Galactic plane found using unsupervised machine learning on the Gaia DR2 observations of the distance and kinematics of stars within 1\,kpc \citep{kounkel2019}.
The fact that the same work identifies that the youngest filaments ($<$\,100\,Myr) are orthogonal to the Local Arm may also be a relevant hint to establish a link between the H{\sc i} filaments and the process of star formation.
However, most of the structures identified in \cite{kounkel2019} are located at intermediate Galactic latitude and are outside of the region of THOR-H{\sc i}. 

\begin{figure}[ht!]
\centerline{\includegraphics[width=0.49\textwidth,angle=0,origin=c]{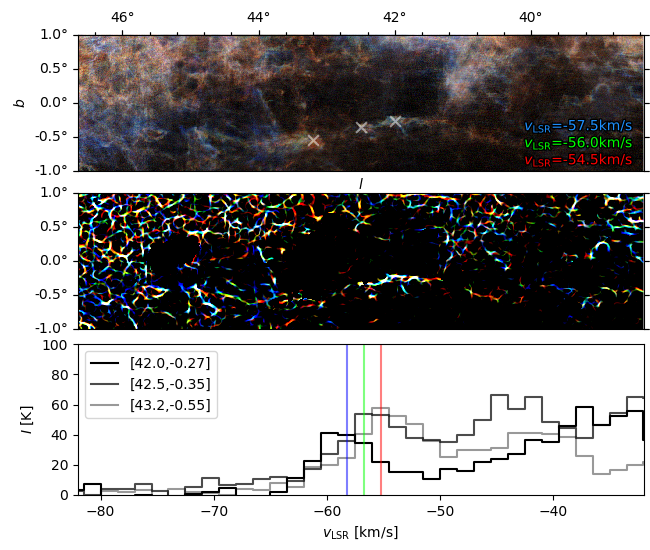}}
\caption{Magdalena H{\sc i} filament.
{\it Top}. Atomic hydrogen (H{\sc i}) emission in three velocity channels.
{\it Middle}. Filaments obtained using the Hessian technique toward the corresponding channels.
{\it Bottom}. Spectra toward the positions indicated by the crosses in the top panel. 
The vertical lines indicate the $v_{\rm LSR}$ that corresponds to the emission shown in the other two panels.}
\label{fig:ExampleMaggie}
\end{figure}

We show another example of a very long filamentary structure that is coherent across velocity channels in Fig.~\ref{fig:ExampleMaggie}. 
Given its uniqueness, we have named it Magdalena, after the longest river in Colombia.
The Magdalena (``Maggie'') filament extends across approximately 4\deg\ in Galactic longitude. 
With a central velocity $v_{\rm LSR}$\,$\approx$\,$-$54\,km\,s$^{-1}$, assuming circular motions, Maggie would be located at approximately 17\,kpc from the Sun and 500\,pc below the Galactic plane with a length that exceeds 1\,kpc.
The linewidths \juan{obtained with a Gaussian decomposition} of the H{\sc i} emission across this filament \juan{are around 13\,\kps\,FWHM \citep{syed2020maggie}}.
\juan{This} indicates that it is mostly likely a density structure, such as those identified in \cite{clark2019}, rather than fluctuations imprinted by the turbulent velocity field \citep[velocity caustic,][]{lazarianANDpogosyan2000,lazarianANDyuen2018}.
The physical processes that would produce such a large structure, if the kinematic distances provide an accurate estimate of its location, is still not well understood but can provide crucial insight into the dynamics of the atomic gas in the Galactic disk and halo.
We present a detailed study of Maggie in an accompanying paper \citep{syed2020maggie}.

\subsection{Regions of interest (ROIs) in the $lv$-diagram}\label{sec:ROIs}

\subsubsection{ROI A: $l$\,$\approx$\,27\deg\ and $v_{\rm LSR}$\,$\approx$\,100\,km\,s$^{-1}$}

We found that the most prominent association of tiles with H{\sc i} filaments perpendicular to the Galactic plane is located around $l$\,$\approx$\,27\deg\ and $v_{\rm LSR}$\,$\approx$\,100\,km\,s$^{-1}$, marked as ROI A in Fig.~\ref{fig:PRSandHIIregions}.
This position has been previously singled out in the study of the H{\sc i} emission toward the Galactic plane, particularly in observations that indicate the presence of voids several kiloparsecs in size centered approximately on the Galactic center, both above and below the Galactic plane \citep{lockman2016}.
These voids, which appear to map the boundaries of the Galactic nuclear wind, are evident in the sharp transition at galactocentric radius of around 2.4\,kpc from the extended neutral gas layer characteristic of much of the Galactic disk to a thin Gaussian layer with 125\,pc FWHM.

\begin{figure*}[ht!]
\centerline{\includegraphics[width=0.99\textwidth,angle=0,origin=c]{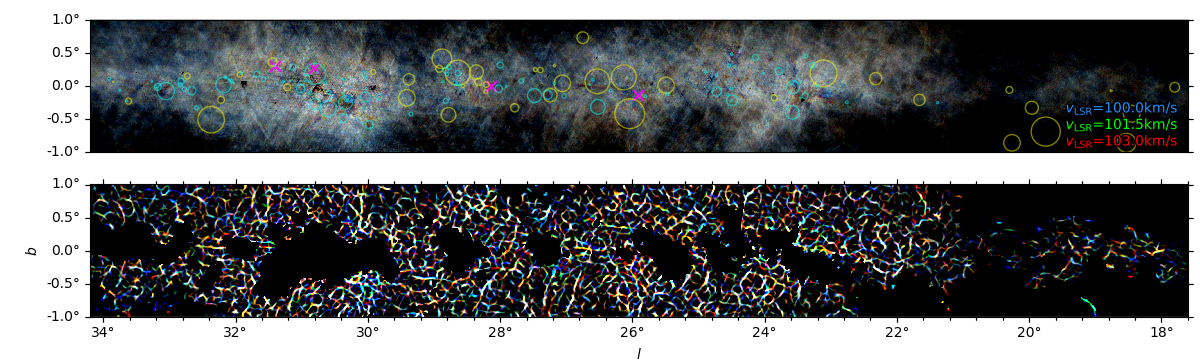}}
\caption{THOR-H{\sc i} (top) observations and their respective filaments from the Hessian analysis (bottom) toward the Galactic longitude range of the region of interest (ROI) A, as defined in Fig~\ref{fig:PRSandHIIregions}.
The yellow circles correspond to the positions and sizes of the supernova remnants in the \cite{green2019} catalogue.
The cyan circles correspond to the positions and sizes of the H{\sc ii} regions in the \cite{anderson2014} catalogue with $v_{\rm LSR}$ within $\pm$10\,\kps\ of the central velocity indicated in the figure.
The magenta crosses mark the position of the THOR OH 1720-MHz masers in that velocity range.}
\label{fig:ROIaRGB}
\end{figure*}

It is plausible that this reported thinning of the H{\sc i} disk is related to the change in the \juan{preferential} orientation of the H{\sc i} filaments at $l$\,$\approx$\,27\deg.
Visual inspection of a few velocity channels within the $l$ and $v_{\rm LSR}$ ranges of the ROI A, shown in Fig.~\ref{fig:ROIaRGB}, indicate that there is indeed a thinning of the Galactic plane for $l$\,$\lesssim$\,22\deg. 
But the H{\sc i} filamentary structures in the range $l$\,$\lesssim$\,22\deg\ and $v_{\rm LSR}$\,$\gtrsim$\,70\,\kps\ are mostly parallel to the Galactic plane, as shown by the positive $V$ values in that range in Fig.~\ref{fig:PRSandHIIregions}.

It is also possible that the filaments perpendicular to the Galactic plane are prominent just because of a geometrical effect.
The tangent point of the Scutum arm is close to $l$\,$\approx$\,30\deg\ \citep{reid2016}, thus, the filaments that are parallel to this spiral arm will appear shortened in the plane of the sky and their orientation will be either random or dominated by the passage of the spiral arm.
However, if that was the case, we should also see a significant variation around $l$\,$\approx$\,50\deg, toward the tangent of the Sagittarius arm.
Such an effect was seen in the strong excess of Faraday rotation toward that position \citep{shanahan2019}, but this is not seen in the orientation of the H{\sc i} filament, as illustrated in Fig.~\ref{fig:PRSandHIIregions}.

A different effect that singles out the position of the ROI A in $l$ and $v_{\rm LSR}$ is the Galactic feature identified as the Molecular Ring around 4\,kpc from the Galactic centre \citep{cohenANDthaddeus1977,roman-duval2010}.
This structure forms two branches with tangent points at $l$\,$\approx$\,25\deg\ and 30\deg\ close to the positive terminal velocity curve, 
which can be reproduced in numerical simulations that include the gravitational potential of the Galactic bar, thus indicating that the presence of this feature does not depend on local ISM heating and cooling processes \citep{fux1999,rodriguez-fernandez2008,sormani2015}.
It is currently unclear whether this structure is really a ring, or emission from many spiral arms crowded together, as suggested by the numerical simulations. 
The coincidence between the tangent points of the Galactic Ring and the ROI A suggests that there is an imprint of the Galactic dynamics in the vertical structure of the atomic hydrogen.
The fact that a similar effect is not observed in the tangent of the Sagittarius arm implies that this is not a geometrical effect or the result of the passage of a spiral arm.

One observational fact that distinguishes ROI A is the large density of H{\sc ii} regions and sources of RRLs, also shown in Fig.~\ref{fig:PRSandHIIregions}.
There is observational evidence of a significant burst of star formation near the tangent of the Scutum arm, in the form of a large density of protoclusters around W43 \citep{motte2003} and multiple red supergiant clusters \citep{figer2006,davies2007,alexander2009}.
This star formation burst can be associated to an enhancement of SN feedback forming a ``Galactic fountain'' \citep{shapiro1976,bregman1980,fraternali2017,kim2018}, whose remants shape the vertical H{\sc i} filaments.

The relation between the star-formation and the vertical H{\sc i} filaments is further suggested by the observed asymmetry in the density of H{\sc i} clouds in the lower Galactic halo toward the tangent points in the first and the fourth quadrants of the Milky Way reported in \cite{ford2010}.
There, the authors show that there are three times more disk-halo clouds in the region $16\pdeg9$\,$<$\,$l$\,$<$\,35\pdeg3 than in the region $324\pdeg7$\,$<$\,$l$\,$<$\,343\pdeg1 and their scale height is twice as large.
Our results indicate that the potential origin of this population of disk-halo clouds, which are found in the range $|b|$\,$<$\,20\deg, \juan{also has} a significant effect in the structure of the H{\sc i} gas at lower Galactic latitude.
Given the symmetry in the Galactic dynamics of the tangent point in the first and the fourth quadrant, the difference between these two regions appear to be linked to the amount of star formation and SN feedback.
This hypothesis is reinforced by the prevalence of vertical H{\sc i} filaments in ROI B, where the effect of the Galactic dynamics is less evident.

\subsubsection{ROI B: $l$\,$\approx$\,36\deg\ and $v_{\rm LSR}$\,$\approx$\,40\,km\,s$^{-1}$}

\begin{figure*}[ht!]
\centerline{\includegraphics[width=0.99\textwidth,angle=0,origin=c]{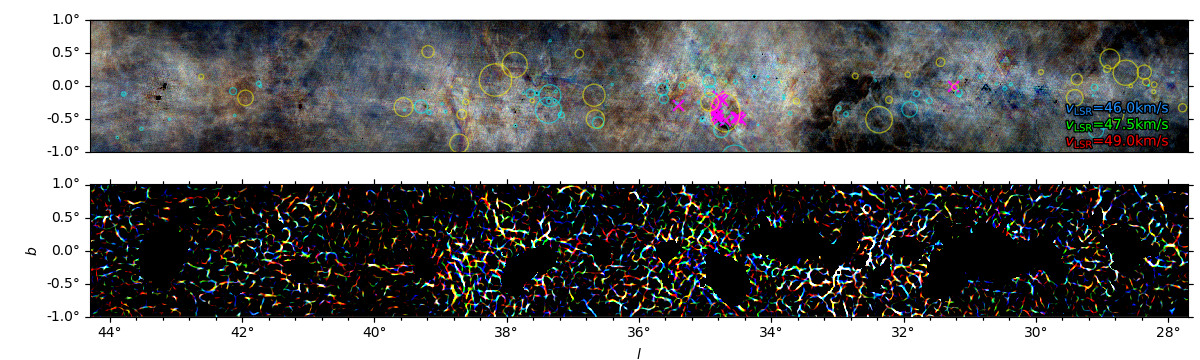}}
\caption{Same as Fig.~\ref{fig:ROIaRGB} for the ROI B.
}
\label{fig:ROIbRGB}
\end{figure*}

The second region where we found a significant association of H{\sc i} filaments perpendicular to the Galactic plane is located around $l$\,$\approx$\,36\deg\ and $v_{\rm LSR}$\,$\approx$\,40\,km\,s$^{-1}$.
Figure~\ref{fig:ROIbRGB} shows that the filamentary structure in the H{\sc i} emission toward ROI B forms an intricate network where many orientations are represented.
However, the values of $V$\,$\ll$\,$0$ and $\left<\theta\right>$\,$\approx$\,90\deg\ indicate that statistically, these filaments are preferentially perpendicular to the Galactic plane.
We also can identify some prominent vertical filaments around $l$\,$\approx$\,38\deg.

The region of interest B coincides with a large density of H{\sc ii} regions, as illustrated in Fig.~\ref{fig:PRSandHIIregions}.
However, this is not enough to establish a causality between the presence of H{\sc ii} regions and a preferential orientation in the H{\sc i} filamentary structure.
As a matter of fact, we did not find a prevalence of vertical H{\sc i} filaments toward W49 ($l$\,$=$\,43\pdeg2, $b$\,$=$\,0\pdeg0, $v_{\rm LSR}$\,$\approx$\,11\,km\,s$^{-1}$) and W51 ($l$\,$=$\,49\pdeg4, $b$\,$=$\,$-$0\pdeg3, $v_{\rm LSR}$\,$\approx$\,60\,km\,s$^{-1}$), two of the most prominent regions of high-mass star formation in the Galaxy \citep{urquhart2014b}.
Both W49 and W51 are relatively young and have a small ratio of SN remnants relative to the number of H{\sc ii} regions \citep{anderson2017}.
The absence of a preferred vertical orientation in the H{\sc i} filament towards these two regions suggests that the effects observed toward ROI A and B are not associated with very young star-forming regions, where most of the massive stars are still on the main sequence.

A plausible explanation for the prevalence of vertical filaments in both ROI A and B is the combined effect of multiple SNe.
In this case, however, it does not correspond to the walls of expanding bubbles, such as those identified in \cite{heiles1984} or \cite{McClure-Griffiths2002}, but rather the preferential orientation produced by multiple generations of bubbles, which depressurize when they reach the scale height and produce structures that are perpendicular to the Galactic plane.
There are at least six 0\pdeg5-scale supernova remnants (SNRs) toward ROI B, including Westerhout 44 \citep[W44,][]{westerhout1958}, as shown in Fig.~\ref{fig:ROIbRGB}.
There is also a strong concentration \juan{of} OH 1720-MHz masers toward ROI B, which are typically excited by shocks and found toward either star-forming regions or SNRs \citep[see for example,][]{frail1994,green1997,wardle2002}. 
Most of these OH 1720-MHz masers appear to be associated with W44 and would not trace the effect of older SNe that may be responsible for the vertical H{\sc i} filamentary structures that are seen, for example, around $l$\,$\approx$\,38\pdeg0 and $l$\,$\approx$\,38\pdeg8 in Fig.\ref{fig:PRSandHIIregions}.
Thus, we resort to numerical experiments to evaluate this effect statistically in Sec.~\ref{sec:mhdsims}.

\subsubsection{ROI C: Riegel-Crutcher cloud}

\begin{figure}[ht!]
\centerline{\includegraphics[width=0.499\textwidth,angle=0,origin=c]{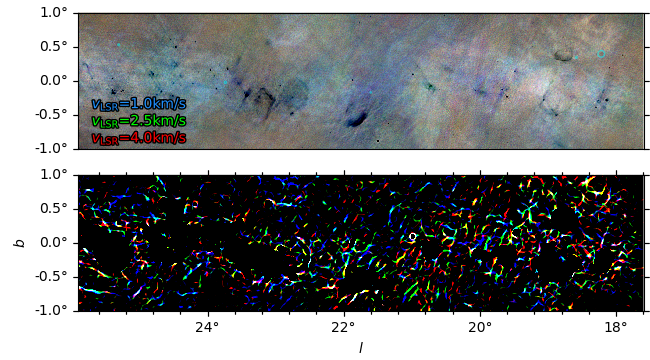}}
\caption{Same as Fig.~\ref{fig:ROIaRGB} for the ROI C.}
\label{fig:ROIcRGB}
\end{figure}

Among the prominent features in the orientation of the H{\sc i} filamentary structure presented in Fig.~\ref{fig:PRSandHIIregions}, one that is of particular interest is that found around 18\deg\,$<$\,$l$\,$<$\,26\deg\ and 0\,$<$\,$v_{\rm LSR}$\,$<$\,10\,km\,s$^{-1}$.
This location corresponds to the Riegel-Crutcher (RC) cloud, a local \citep[$d$\,$=$\,125\,$\pm$\,25\,pc,][]{crutcherANDlien1984} CNM structure that extends approximately 40\deg\ in Galactic longitude and 10\deg\ in latitude \citep{riegelANDjennings1969,crutcherANDriegel1974}.
Figure~\ref{fig:ROIcRGB} reveals features in the emission around $l$\,$\approx$\,22\deg\ that are reminiscent of the strands found in the most studied portion of the RC cloud \citep[$|l|$\,$<$\,5\deg\ and $|b|$\,$<$\,5\deg,][]{McClure-Griffiths2006,denes2018}, but with a different orientation.
The analysis of the filamentary structure presented in this paper serves as an alternative way of defining the RC cloud, through the contrast \juan{in the orientation of the filamentary structures with respect to that in the surrounding velocity channels}.

Many of the structures in the RC cloud are seen as shadows against an emission background.
It is common at low Galactic latitudes that cold foreground H{\sc i} clouds absorb the emission from the H{\sc i} gas behind. 
This effect is often called H{\sc i} self-absorption (HISA), although it is not self-absorption in the standard radiative transfer sense, because the absorbing cloud may be spatially distant from the background H{\sc i} emission, but \juan{share} a common radial velocity \citep{knapp1974,gibson2000}.

\begin{figure}[ht!]
\centerline{\includegraphics[width=0.499\textwidth,angle=0,origin=c]{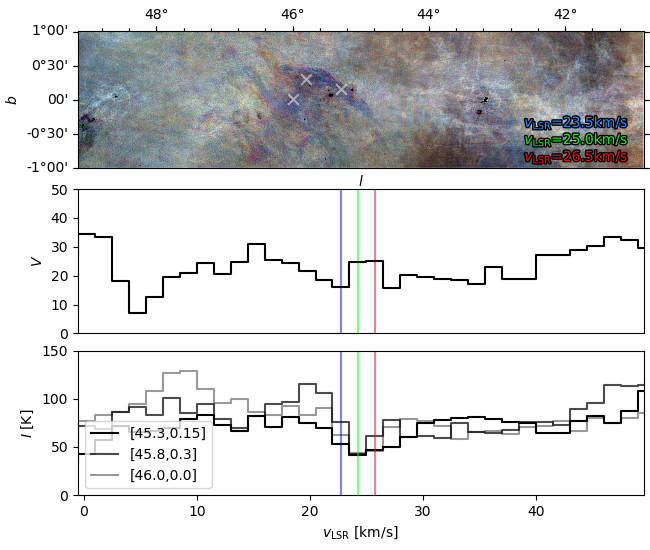}}
\caption{
Selected example of a prominent H{\sc i} self-absorption (HISA) structure.
{\it Top}. Atomic hydrogen (H{\sc i}) emission in three velocity channels.
{\it Middle}. Projected Rayleigh statistic ($V$, Eq.~\ref{eq:myprs}) corresponding to the orientation of the filamentary structures in the H{\sc i} emission in the corresponding channels.
{\it Bottom}. Spectra toward the positions indicated by the crosses in the top panel. 
The vertical lines indicate the $v_{\rm LSR}$ that corresponds to the emission shown in the other two panels.}
\label{fig:ExampleHISA}
\end{figure}

HISAs are often identified by a narrow depression in the H{\sc i} spectrum and as a coherent shadow in the emission maps.
The systematic identification of HISA features and the evaluation of the completeness of the census of cold H{\sc i} that it provides is a complex process, as described in \cite{gibson2005b} and \cite{kavars2005} or \cite{wang2020hisa} and \cite{syed2020hisa} in the particular case of THOR-H{\sc i}.
The H{\sc i} filament orientations provide a complementary method to identify HISA structures, by quantifying the difference in orientation of the HISA with respect to that of the intensity background.
Figure~\ref{fig:ExampleHISA} shows an example of the HISA identification in the spectrum and in the orientation of the filamentary structure toward the molecular cloud GRSMC 45.6+0.3 studied in \cite{jackson2002}.
The spectra and the values of $V$ across velocity channels toward this region indicate that the HISA feature is evident in both the intensity and the morphology of the H{\sc i} emission.

\subsubsection{ROI D: Terminal velocities}

The final exception to the general trend of H{\sc i} filaments parallel to the Galactic plane is found at the maximum and minimum $v_{\rm LSR}$ with a large prominence in multiple velocity channels at $l$\,$>$\,56\deg.
Part of the emission at the extremes of the radial velocities has been identified as a separate H{\sc i} component from that in the Galactic disk and belongs to the Galactic Halo \citep{shane1971,lockman2002}.
This component is 1 to 2\,kpc above the Galactic plane, and it is usually called ``extraplanar'' H{\sc i} gas, which avoids potential confusion with the Galactic Halo as intended in, for example, cosmological simulations \citep[see for example,][]{fraternaliANDbinney2006,fraternaliANDbinney2008,fraternali2017}.

\begin{figure*}[ht!]
\centerline{
\includegraphics[width=0.33\textwidth,angle=0,origin=c]{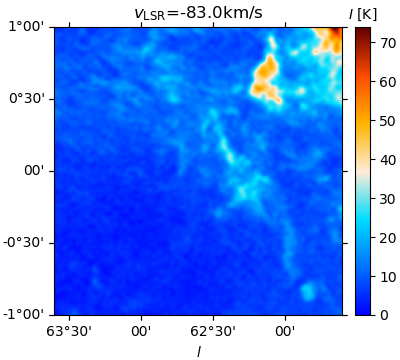}
\includegraphics[width=0.33\textwidth,angle=0,origin=c]{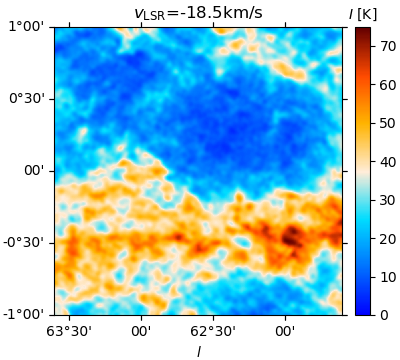}
\includegraphics[width=0.33\textwidth,angle=0,origin=c]{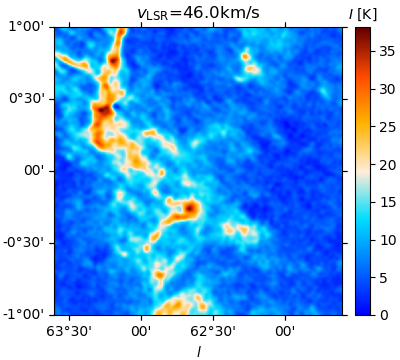}
}
\caption{THOR-H{\sc i} emission for two velocity channels in the region of interest (ROI) D (left and right) and a channel toward the same $l$ range but dominated by filaments parallel to the Galactic plane (middle).}
\label{fig:ROIdRGB}
\end{figure*}

Figure~\ref{fig:ROIdRGB} shows that the structure of the emission at $-83$ and $46$\,km\,s$^{-1}$ is clearly vertical and different from that at intermediate velocities.
It is also different from the structure in ROIs A and B, shown in Figs.~\ref{fig:ROIaRGB} and \ref{fig:ROIbRGB}, where the vertical H{\sc i} structures are part of an intricate network of filamentary structures.
In the ROI D, most of the vertical filamentary structures in H{\sc i} emission are clearly separated from other structures and \juan{notably} extend to high $b$, further suggesting that they are H{\sc i} clouds from the halo.
At least one of the tiles with preferentially vertical filaments corresponds to a cloud above the maximum velocity allowed by Galactic rotation, that at $l$\,$=$\,60\pdeg8 and $v_{\rm LSR}$\,$=$\,58.0\,\kps\ identified in the VGPS \citep{stil2006b}. 

\cite{marascoANDfraternali2011} studied the extraplanar H{\sc i} gas of the Milky Way, and found that it is associated with SN feedback and mainly consists of gas that is falling back to the MW after being ejected by SNe.
The reason why the extraplanar H{\sc i} is more conspicuous falling down than going up is because it cools while settling down, whereas it is hotter and so less visible when going up, as can be seen in the numerical simulations presented in \cite{kim2015b} and \cite{girichidis2016}.
So potentially, the vertical filaments around ROI D are falling back after being ejected by SNe instead of material going up, as is possibly the case for ROI A and ROI B.
The time and spatial delay between SN explosion and gas falling back might also explain why we do not find that every region associated with SNe has vertical filaments.

\subsection{Relation to the structure of the molecular gas}

Filamentary structures designated giant molecular filaments (GMF) have been previously identified toward the Galactic plane in the emission from molecular species, such as $^{13}$CO \citep[][]{goodman2014,ragan2014,wang2015,zucker2015,abreu-vicente2017,wang2020gmf}.
To establish a link between the H{\sc i} filaments and the GMFs, without detailing individual objects, we also applied the Hessian analysis to the $^{13}$CO\,($J$\,$=$\,1\,$\rightarrow$\,0) emission observations in the GRS survey.
Following the selection criteria presented in Sec.~\ref{sec:filselection}, we estimated the orientation of the filamentary structures identified using the Hessian method in the GRS observations projected into the same spectral axis of THOR-H{\sc i}.

\begin{figure}[ht!]
\centerline{
\includegraphics[width=0.495\textwidth,angle=0,origin=c]{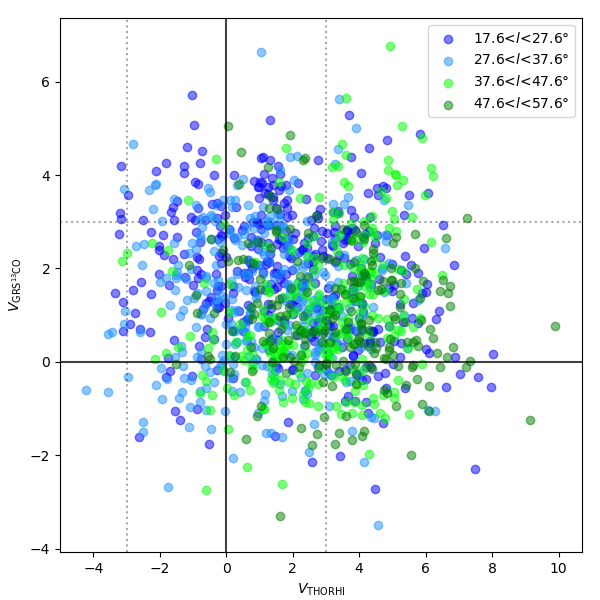}
}
\caption{Comparison between the values of the projected Rayleigh statistic $V$ calculated using the THOR and $^{13}$CO GRS observations in the indicated Galactic-longitude ranges.
Both are calculated using a derivative kernel with 120\arcsec\ FWHM.
Each point in the scatter plot corresponds to the values estimated using the Hessian analysis in the same 2\deg\,$\times$\,2\deg\ tile and 1.5-\kps\ velocity channel in both datasets.}
\label{fig:scatterTHORandGRS_PRS}
\end{figure}

Figure~\ref{fig:scatterTHORandGRS_PRS} shows a comparison of the values of $V$ in THOR-H{\sc i} and GRS observations.
In agreement with the GMF compilation presented in \cite{zucker2018}, we find that most of the $^{13}$CO filamentary structures are parallel to the Galactic plane.
However, 
we found no evident correlation between the orientation of the filamentary structures in \juan{the two} tracers.

There can be several reasons for this general lack of correlation between the H{\sc i} and the $^{13}$CO filamentary structures.
First, the linewidths of the $^{13}$CO emission are narrower than those of the H{\sc i} and it is possible that we are washing away part of the orientation of the filaments by projecting both data sets into the same spectral grid.
This effect, however, may not be dominant.
The Gaussian decomposition of the GRS presented in \cite{riener2020} indicates that the mean velocity dispersion ($\sigma_{v}$) is approximately 0.6\,km\,s$^{-1}$ and the interquartile range is 0.68\,$<$\,$\sigma_{v}$\,$<$\,1.89\,km\,s$^{-1}$, thus re-gridding the data to 1.5\,km\,s$^{-1}$ resolution would not completely alter the morphology of most of the emission.
The fact that we found a large number of emission tiles with $V$\,$\gg$\,\juan{0} indicates that there is a preferential orientation of the filamentary structures in $^{13}$CO, parallel to the Galactic plane, and this preferential orientation is not washed away by the integration over a broad spectral range.

Second, although there is a morphological correlation of the H{\sc i} and the $^{13}$CO, as quantified in \cite{soler2019}, the filamentary structure in the H{\sc i} emission is not necessarily related to that of the $^{13}$CO.
In general, the much larger filling factor of the H{\sc i} makes it unlikely that most of its structure is related to that of the less-filling molecular gas.
Moreover, when evaluating comparable scales, the H{\sc i} and the $^{13}$CO can appear completely decoupled \citep[][]{beuther2020}.
This does not discard the local correlation between the morphology of \juan{the two} tracers toward filamentary structures, as reported in \cite{wang2020hisa} and \cite{syed2020hisa}, but it shows that the orientation of the filamentary structures are generally different. 

\section{Comparison to MHD simulations}\label{sec:mhdsims}

Among the plethora of physical processes that can be responsible for the preferential orientation of H{\sc i} filamentary structures reported in this paper, we \juan{explored} the effect of SNe, magnetic fields, and galactic rotation in the multiphase medium in two families of \juan{state-of-the-art and readily-available} numerical simulations.
First, we \juan{considered} the simulations in the ``From intermediate galactic scales to self-gravitating cores'' (FRIGG) project, introduced in \cite{hennebelle2018}, which we used to explore the effects of SN feedback and magnetization in the orientation of the H{\sc i} structures.
Second, we \juan{considered} the {\tt Cloud Factory} simulation suite, which is introduced in \cite{smith2020} and is designed to study SN feedback effects while also including the larger-scale galactic context in the form of the galactic potential, the differential rotation of the disk, and the influence of spiral arms.


\subsection{Supernovae and magnetic fields}\label{sec:mhdsimsFRIGG}

\subsubsection{Initial conditions}

The FRIGG simulations \juan{used} the {\tt RAMSES} code and take place in a stratified 1-kpc side box with SN explosions and MHD turbulence.
This is a standard configuration that can be found in other works, such as, \cite{deavillez2007}, \cite{walch2015}, and \cite{kim2017}.
It includes the cooling and heating processes relevant to the ISM, which produce a multiphase medium.
They reproduce the vertical structure of the Galactic disk, which results from the dynamical equilibrium between the energy injected by the SNe and the gravitational potential of the disk.
In particular, we \juan{used} the set of simulations described in \cite{iffrig2017}, which have different levels of magnetization that we analyze to assess the role of the magnetic field in the orientation and characteristics of the H{\sc i} filaments.
These simulations have a resolution that is limited to 1.95\,pc, however, this is enough for a first glimpse at the orientation of the structures formed under the general 1-kpc scale initial conditions.

\cite{iffrig2017} report that the efficiency of the SNe in driving the turbulence in the disk is rather low, of the order of 1.5\%, and strong magnetic fields increase it by a factor of between two and three.
It also reports a significant difference introduced by magnetization in the filamentary structures perpendicular to the Galactic plane, illustrated in their figure~1.
To quantify the differences introduced by the magnetization in the morphology of the emission from atomic hydrogen, we compared one snapshot in the simulation with ``standard'' magnetization, initial magnetic field $B_{0}$ of about 3\,$\mu$G, and one with ``very high" magnetization, $B_{0}$\,$\approx$\,12\,$\mu$G.
Both simulations have an initial particle density of $n_{0}$\,$=$\,1.5\,cm$^{-1}$.
The initial magnetic field strengths are chosen around the median magnetic field strength 6.0\,$\pm$\,1.8\,$\mu$G observed in the CNM \citep{heilesANDtroland2005}.
We selected snapshots at 75 and 81\,Myr for the standard and very high magnetization cases, respectively, both of which are available in the simulation database {\tt Galactica} (\href{http://www.galactica-simulations.eu}{http://www.galactica-simulations.eu}).
This selection guarantees that the simulations have reached a quasi-steady state and does not affect the reported results.
\juan{We present the synthetic observations in Fig.~\ref{fig:FRIGGrgb} and describe details of their construction in App.~\ref{app:FRIGG}.} 

\subsubsection{Hessian analysis results}

The prevalence of longer filamentary structure with higher magnetization has been reported in previous studies  \citep{hennebelle2013a,seifriedANDwalch2015,solerANDhennebelle2017}.
It is related to the effect of strain, which means that these structures simply result from the stretch induced by turbulence, and the confinement by the Lorentz force, which therefore leads them to survive longer in magnetized flows.
However, their orientation in this kind of numerical setup has not been systematically characterized.

\juan{We applied the Hessian analysis with a 120\arcs\ FWHM derivative kernel to the synthetic H{\sc i} emission from the FRIGG simulations. 
We summarized the results in Fig.~\ref{fig:FRIGGresults}}.
Our first significant finding is that the standard magnetization case reproduces some of the filamentary structures parallel to the Galactic plane that are broadly found in the observations, but these do not show the same significance in terms of the values of $V$.
This means that the initial conditions in the standard case reproduce some of the horizontal filaments just with the anisotropy introduced by the vertical gravitational potential.
However, this setup is missing the Galactic dynamics that are the most likely source of the stretching of structures in the direction of the plane.

Our second finding is that the magnetization does not constrain the filamentary structures to the plane, but rather maintains the coherence of the structures that are blown in the vertical direction by the SNe, as show in the high magnetization case.
When the bubble blown by a SN reaches the scale height and depressurizes, the magnetic field maintains the coherence in its walls, which would fragment if the magnetic field were weaker.
Subsequently, the gas layer consisting of ejected clouds falls back on the plane and is stretched along the field lines.

The aforementioned results suggest that the magnetic field may play a significant role in the prevalence of vertical structures in the regions indicated in Fig.~\ref{fig:PRSandHIIregions}.
Filamentary structures have been observed in radio continuum towards the Galactic center and their radio polarization angles indicate that these structures follow their local magnetic field \citep[see for example,][]{morris1996,yusef-zadeh2004}.
Studies of radio polarization at higher Galactic latitude indicate a correspondence between the depolarization canals and the H{\sc i} filamentary structures, which suggests that the filamentary structures share the orientation of the magnetic field \citep{kalberla2017}.
Thus, it is tempting to think that a similar effect can be responsible for the orientation of the vertical filaments in the \thorhi\ observations.

\subsection{Supernovae and Galactic dynamics}\label{sec:mhdsimsAREPO}

\subsubsection{Initial conditions}

We \juan{considered} the effect of the Galactic dynamics by using the {\tt CloudFactory} simulations, presented in \cite{smith2020}.
These simulations, \juan{ran} using the {\tt AREPO} code, consist of a gas disk inspired by the Milky Way gas disk model of \cite{McMillan2017} and focus on the region between 4 and 12\,kpc in galactocentric radius.
The simulations start with a density distribution of atomic hydrogen that declines exponentially at large radii.
Molecular hydrogen forms self-consistently as the gas disk evolves.
We used a 1-kpc-side box region within the large-scale setup, with a mass resolution of $10$\,$M_{\odot}$, and gas self-gravity.

We compared two simulation setups.
In one, SN were placed randomly in the galactic disk at a fixed rate of 1 per 50 years, chosen to match the value appropriate for the Milky Way.
The other setup combined a random SN component with a much smaller rate of 1 per 300 years, designed to represent the effect of type Ia SNe, with a clustered SN component whose rate and location were directly tied to the rate and location of star formation in the simulation.
Following the terminology of \cite{smith2020}, we refer to these simulations as potential-dominated and feedback-dominated, respectively.

\cite{smith2020} reports the alignment of filamentary structures in the disk by spiral arms and the effect of differential rotation.
The authors also note that clustered SN feedback randomize the orientation of filaments and produce molecular cloud complexes with fewer star-forming cores.
To quantify these effects in the H{\sc i} emission, we studied the synthetic observation of one snapshot in the potential- and feedback-dominated simulations.
\juan{We present the synthetic observations in Fig.~\ref{fig:CloudFactoryRGB} and describe details on their construction in App.~\ref{app:CloudFactory}.} 

\juan{The simulations of the whole disc were performed at a lower resolution with cell masses of 1000\,M$_\odot$ for 150\,Myr until the simulated ISM reached a steady state. 
This initial setup step was followed by a progressive resolution increase in a corotating 1-kpc sized box. 
The snapshot used for analysis is chosen when the gas within the high resolution box has had time to perform two spiral arm passages, enough for the simulated ISM to adapt to the new resolution and physics.
Unfortunately, the simulations did not progress much further in time such that an analysis of the simulation at different ISM conditions is not possible at the moment.}

\subsubsection{Hessian analysis results}

\juan{We applied the Hessian analysis with a 120\arcs\ FWHM derivative kernel to the synthetic H{\sc i} emission from the {\tt CloudFactory} simulations. 
The results are summarized in Fig.~\ref{fig:CloudFactoryResults}}.
The most significant outcome of this study is that the Galactic dynamics in these simulations naturally produce filamentary structures parallel to the Galactic plane across velocity channels, which are comparable to those found in the GALFA-H{\sc i} and THOR-H{\sc i} observations.
These filamentary structures are coherent across several velocity channels and correspond to overdensities that are clearly identifiable in the density cubes from the simulation, thus, they are not exclusively the product of fluctuations in the velocity field.

The clustered SNe in the feedback-dominated simulation produce structures that resemble clumpy filaments in the synthetic H{\sc i} PPV cube, as shown in Fig.~\ref{fig:CloudFactoryRGB}.
These structures do not show a significant preferential orientation, as illustrated by the values of $|V|$\,$\lesssim$\,\juan{0} in the corresponding panel of Fig.~\ref{fig:CloudFactoryResults}.
This confirms and quantifies the randomization of the structures described in \cite{smith2020}.

Both the potential-dominated and feedback-dominated cases considered in this numerical experiment correspond to extreme cases.
The fact that the potential-dominated simulation does not produce a significant number of vertical H{\sc i} filaments suggests that these are most likely related to the effect of clustered SNe.
The fact that the SN feedback erases all the anisotropy introduced by the dynamics in the direction of the Galactic plane indicates that the prevalence of vertical filaments is an indication that, at least in a few specific locations, SN feedback has a dominant effect in the structure of the ISM.
Therefore, the observation of this vertical H{\sc i} filaments is a promising path towards quantifying the effect of SN feedback in the Galactic plane.

\section{Conclusions}\label{sec:conclusions}

We presented a study of the filamentary structure in the maps of the H{\sc i} emission toward inner Galaxy using the 40\arcsec-resolution observations in the THOR survey.
We identified filamentary structures in individual velocity channels using the Hessian matrix method and characterized their orientation using tools from circular statistics.
We analyzed the emission maps in 2\deg\,$\times$\,2\deg\ tiles in 1.5-\kps\ velocity channels to report the general trends in orientation across Galactic longitude and radial velocity.

We found that the majority of the filamentary structures are aligned with the Galactic plane.
This trend is in general persistent across velocity channels.
Comparison with the numerical simulation of the Galactic dynamics and chemistry in the {\tt CloudFactory} project indicate that elongated and non-self-gravitating structures naturally arise from the galactic dynamics and are identified in the emission from atomic hydrogen.

Two significant exceptions to this general trend of H{\sc i} filaments being parallel to the Galactic plane are grouped around $l$\,$\approx$\,37\deg\ and $v_{\rm LSR}$\,$\approx$\,50\,km\,s$^{-1}$ and toward $l$\,$\approx$\,36\deg\ and $v_{\rm LSR}$\,$\approx$\,40\,km\,s$^{-1}$.
They correspond to H{\sc i} filaments that are mostly perpendicular to the Galactic plane.
The first location corresponds to the tangent point of the Scutum arm and the terminal velocities of the Molecular Ring, where there is a significant accumulation of H{\sc ii} regions.
The second position also shows a significant accumulation of H{\sc ii} regions and supernova remnants.

Comparison with numerical simulations in the {\tt CloudFactory} and FRIGG projects indicate that the prevalence of filamentary structures perpendicular to the Galactic plane can be the result of the combined effect of SN feedback and magnetic fields.
These structures do not correspond to the relatively young ($<$\,10\,Myrs) structures that can be identified as shells in the H{\sc i} emission, but rather to the cumulative effect of older SNe that lift material and magnetic fields in the vertical direction.
Thus, their prevalence in the indicated regions \juan{is the potential signature of an even earlier} history of star formation and stellar feedback in the current structure of the atomic gas in the Galactic plane.
\juan{Our observational results motivate the detailed study of more snapshots and different numerical simulations to firmly establish the relation between these relics in the atomic gas and the record of star formation}.

Another exception to the general trend of H{\sc i} filaments being parallel to the Galactic plane is found around the positive and negative terminal velocities.
Comparison with previous observations suggests that these structures may correspond to extraplanar H{\sc i} clouds between the disk and the halo of the Milky Way.
A global explanation for the vertical H{\sc i} filaments is that the combined effect of multiple SNe creates a layer of gas consisting of ejected clouds, some of which are falling back on the plane.
Such clouds would naturally tend to be vertically elongated and coherent due to the effect of the magnetic fields.
Galactic dynamics may be responsible for creating the observed vertical filaments only in an indirect way: it helps bringing the gas together, creating favourable conditions for SNe to cluster together, explode, and create the vertical structure.

The statistical nature of our study unveils general trends in the structure of the atomic gas in the Galaxy and motivates additional high-resolution observations of the H{\sc i} emission in other regions of the Galaxy.
Further studies of the nature and the origin of the H{\sc i} filamentary structures call for the identification of other relevant characteristics, such as their width and length, as well as the physical properties that can be derived using other complementary ISM tracers.
Our results demonstrate that measuring the orientation of filamentary structures in the Galactic plane is a \juan{promising} tool to reveal the imprint of the Galactic dynamics, stellar feedback, and magnetic fields in the observed structure of the Milky Way and other galaxies.

\begin{figure*}[ht!]
\centerline{
\includegraphics[width=0.44\textwidth,angle=0,origin=c]{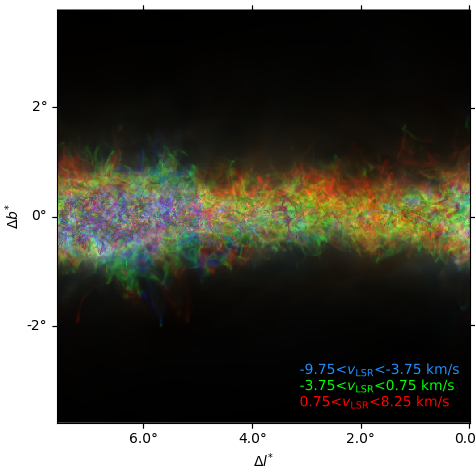}
\includegraphics[width=0.44\textwidth,angle=0,origin=c]{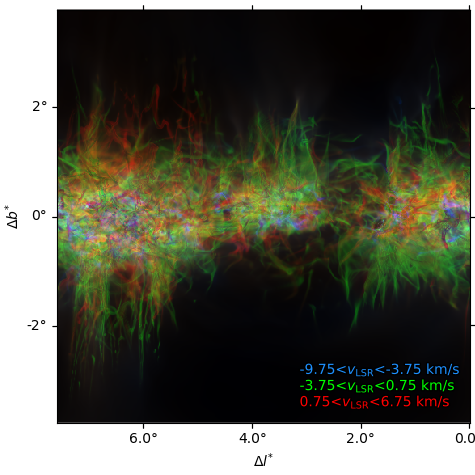}
}
\caption{
Synthetic observations of the H{\sc i} emission from the FRIGG simulations with a standard (3\,$\mu$G) and very high (12\,$\mu$G) initial magnetizations, shown in the left and right panels, respectively.
The colors represent the emission in the three indicated radial velocity bins with the same average emission.
}\label{fig:FRIGGrgb}
\end{figure*}
\begin{figure*}[ht!]
\centerline{
\includegraphics[width=0.49\textwidth,angle=0,origin=c]{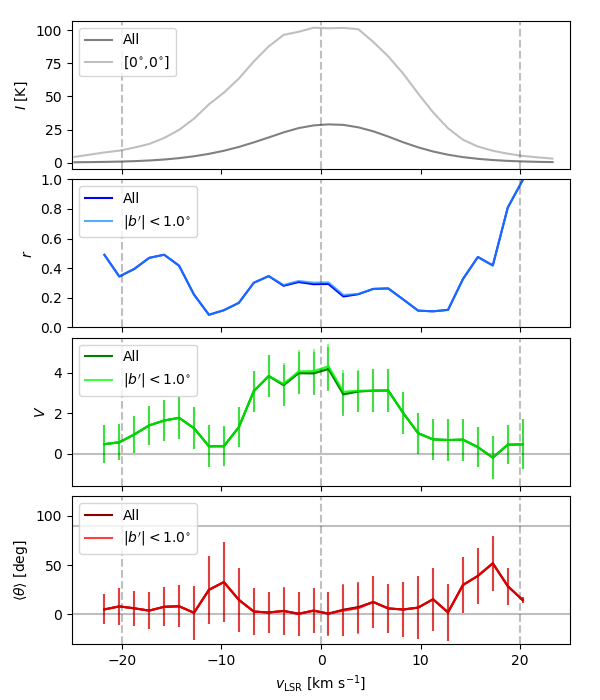}
\includegraphics[width=0.49\textwidth,angle=0,origin=c]{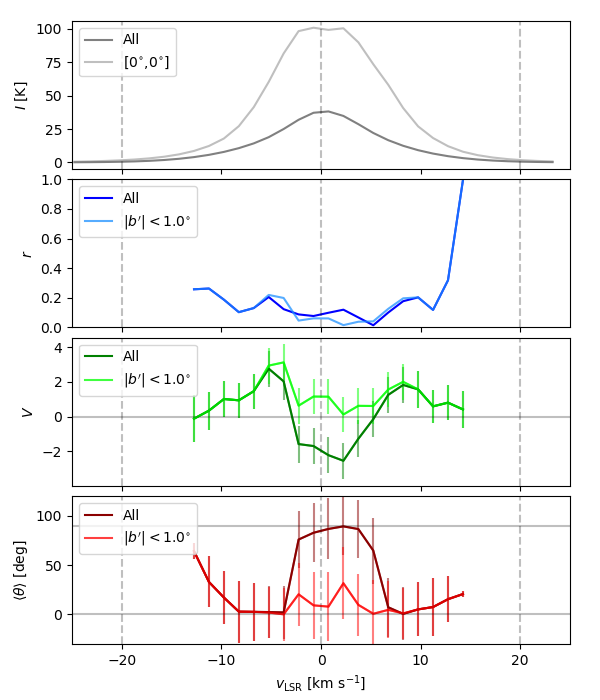}
}
\caption{\juan{Synthetic H{\sc i} emission intensity and} circular statistics used for the study of the orientation of the H{\sc i} filamentary structures across velocity channels in the synthetic observations presented in Fig.~\ref{fig:FRIGGrgb}.
{\it Top}. \juan{H{\sc i} emission averaged over the whole map and toward the indicated position.}
{\it Middle top}. Mean resulting vector ($r$), which indicates if the distribution of orientation angles is flat ($r$\,$\approx$\,0) or sharply unimodal ($r$\,$\approx$\,1).
{\it Middle bottom}. Projected Rayleigh statistic ($V$), which indicates if the distribution of orientation angles is clearly peaked around 0\deg\ ($V$\,$\gg$\,\juan{0}) or 90\deg\ ($V$\,$\ll$\,\juan{0}).
{\it Bottom}. Mean orientation angle ($\left<\theta\right>$) of the H{\sc i} filamentary structures. 
}
\label{fig:FRIGGresults}
\end{figure*}

\begin{figure*}[ht!]
\centerline{
\includegraphics[width=0.44\textwidth,angle=0,origin=c]{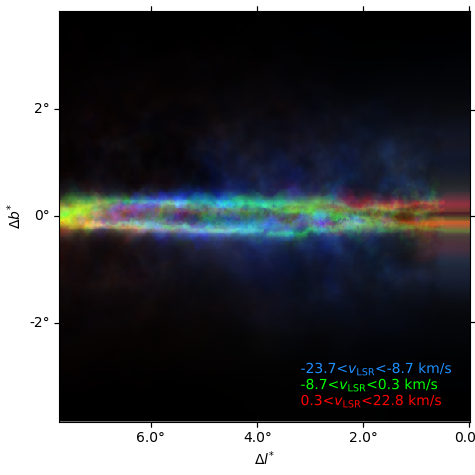}
\includegraphics[width=0.44\textwidth,angle=0,origin=c]{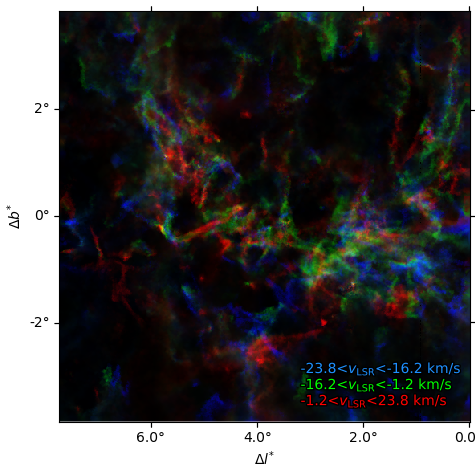}
}
\caption{
Synthetic observations of the H{\sc i} emission from the potential and feedback dominated {\tt CloudFactory} simulations, presented in the left and right panels, respectively.
The colors represent the emission in the three indicated radial velocity bins, each of the with the same average emission.
}\label{fig:CloudFactoryRGB}
\end{figure*}
\begin{figure*}[ht!]
\centerline{
\includegraphics[width=0.49\textwidth,angle=0,origin=c]{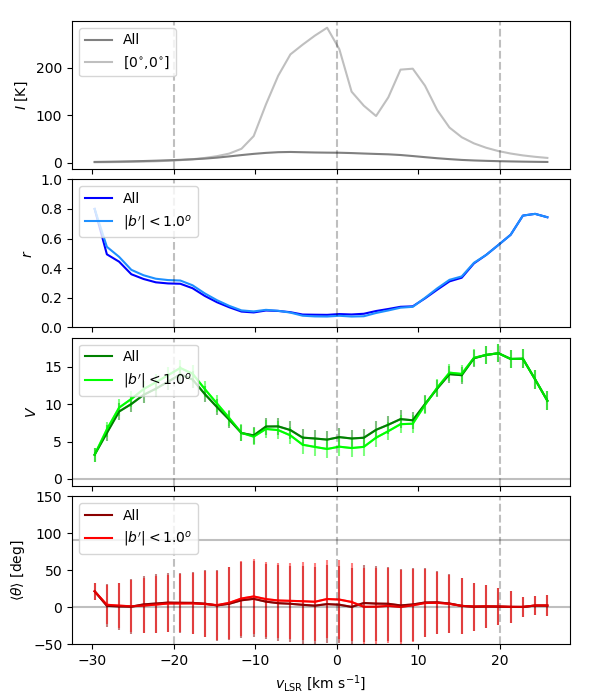}
\includegraphics[width=0.49\textwidth,angle=0,origin=c]{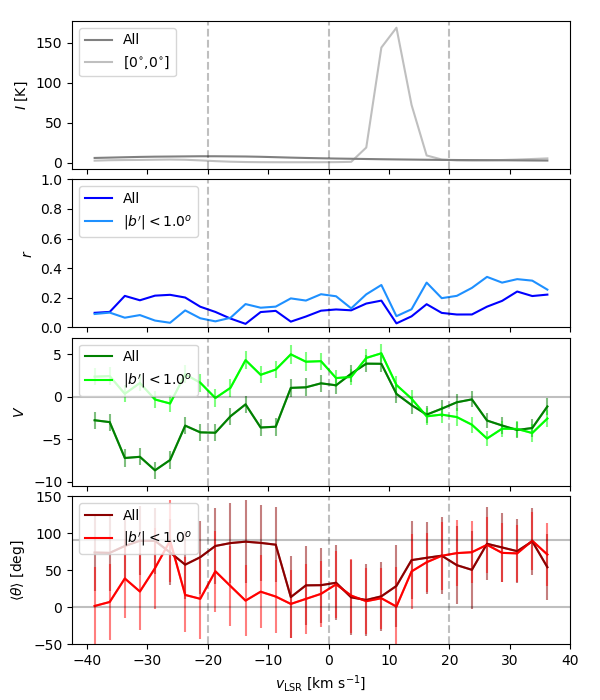}
}
\caption{\juan{Synthetic H{\sc i} emission intensity and} circular statistics used for the study of the orientation of the H{\sc i} filamentary structures across velocity channels in the synthetic observations presented in Fig.~\ref{fig:CloudFactoryRGB}.
{\it Top}. \juan{H{\sc i} emission averaged over the whole map and toward the indicated position.}
{\it Middle top}. Mean resulting vector ($r$), which indicates if the distribution of orientation angles is flat ($r$\,$\approx$\,0) or sharply unimodal ($r$\,$\approx$\,1).
{\it Middle bottom}. Projected Rayleigh statistic ($V$), which indicates if the distribution of orientation angles is clearly peaked around 0\deg\ ($V$\,$\gg$\,\juan{0}) or 90\deg\ ($V$\,$\ll$\,\juan{0}).
{\it Bottom}. Mean orientation angle ($\left<\theta\right>$) of the H{\sc i} filamentary structures.
}\label{fig:CloudFactoryResults}
\end{figure*}

\begin{acknowledgements}
JDS, HB, and YW acknowledge funding from the European Research Council under the Horizon 2020 Framework Program via the ERC Consolidator Grant CSF-648505.
HB, JS, SCOG, and RK acknowledge support from the Deutsche Forschungsgemeinschaft via SFB 881, ``The Milky Way System'' (subprojects A1, B1, B2, and B8).
SCOG and RK also acknowledge support from the Deutsche Forschungsgemeinschaft via Germany's Excellence Strategy EXC-2181/1-390900948 (the Heidelberg STRUCTURES Cluster of Excellence).
RK also acknowledges funding from the European Research Council via the ERC Synergy Grant ECOGAL (grant 855130) and the access to the data storage service SDS@hd and computing services bwHPC supported by the Ministry of Science, Research and the Arts Baden-Württemberg (MWK) and the German Research Foundation (DFG) through grants INST 35/1314-1 FUGG as well as INST 35/1134-1 FUGG.
NS acknowledges support by the Agence Nationale de la Recherche (ANR) and the DFG through the project ``GENESIS'' (ANR-16-CE92-0035-01/DFG1591/2-1). 

We thank the anonymous referee for the thorough review. 
We highly appreciate the comments, which significantly contributed to improving the quality of this paper.
JDS thanks the following people who helped with their encouragement and conversation: Peter G. Martin, Marc-Antoine Miville-Desch\^{e}nes, Josh Peek, Jonathan Henshaw, Steve Longmore, Paul Goldsmith, and Hans-Walter Rix.
Part of the crucial discussions that lead to this work took part under the program Milky-Way-Gaia of the PSI2 project funded by the IDEX Paris-Saclay, ANR-11-IDEX-0003-02. 

This work has been written during a moment of strain for the world and its inhabitants. 
It would not have been possible without the effort of thousands of workers facing the COVID-19 emergency around the globe.
Our deepest gratitude to all of them.

{\it Software}: {\tt matplotlib} \citep{sw:matplotlib2007}, {\tt astropy} \citep{sw:astropy2013}, {\tt magnetar} \citep{sw:soler2020}, {\tt RHT} \citep{sw:clark2020}.
\end{acknowledgements}

\bibliographystyle{aa}
\bibliography{AA_2020_38882.bbl}

\appendix 


\section{Details of the Hessian method implementation}\label{app:Hessian}

\begin{figure*}[ht!]
\centerline{
\includegraphics[width=0.95\textwidth,angle=0,origin=c]{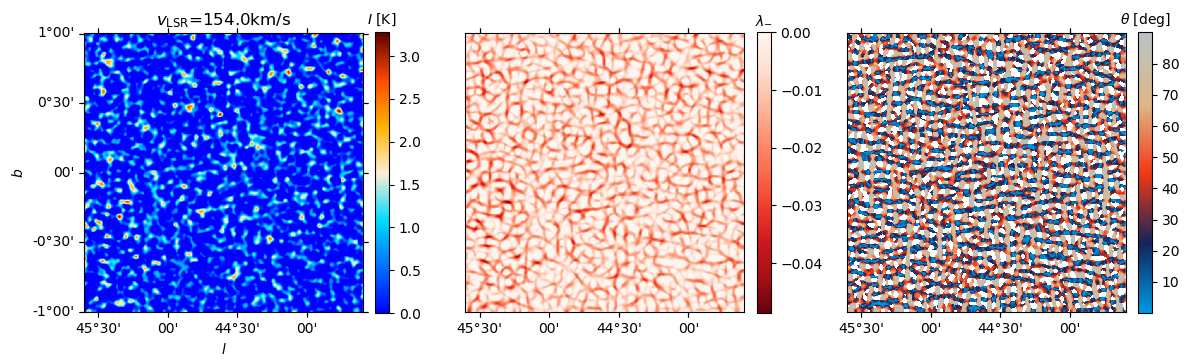}
}
\vspace{-0.3cm}
\caption{
Hessian analysis applied to a velocity-channel map dominated by noise toward the same region presented in Fig.~\ref{fig:example}.
{\it Left}. H{\sc i} intensity map.
{\it Center}. Map of the eigenvalue of the Hessian matrix identified as $\lambda_{-}$ in Eq.~\eqref{eq:lambda}, which is used to characterize the filamentary structures in the intensity map.
The overlaid grey map shows the filamentary structure obtained from the Hessian matrix analysis of the continuum noise maps, see App.~\ref{app:Hessian} for details.
{\it Right}. Map of the orientation angles evaluated using Eq.~\eqref{eq:theta}.
}\label{fig:NoiseChannel}
\end{figure*}

\begin{figure*}[ht!]
\centerline{
\includegraphics[width=0.95\textwidth,angle=0,origin=c]{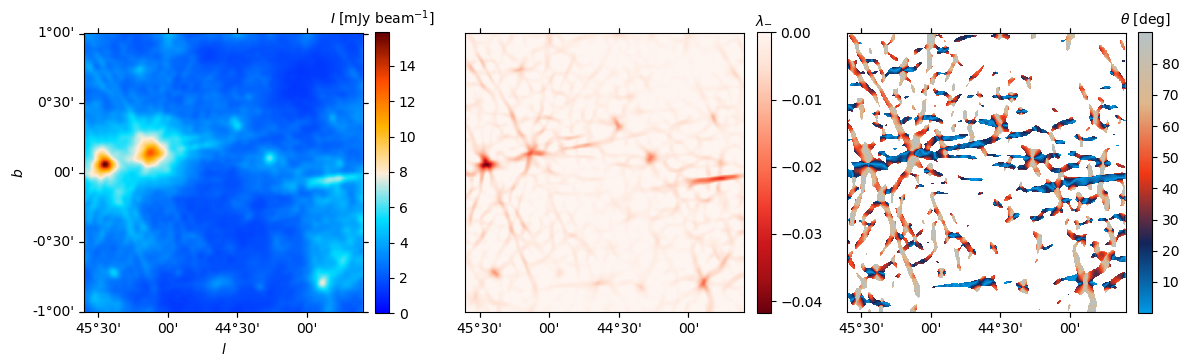}
}
\caption{
Same as Fig.~\ref{fig:NoiseChannel} for the noise map of the continuum emission at 1.42\,GHz ($\sigma_{I_{1.42}}$), which we use as a proxy of the structure introduced by the side-lobes in the imaging of the H{\sc i} emission.
}\label{fig:ContinuumNoise}
\end{figure*}

\subsection{Accounting for the effect of noise}\label{app:MCs}

\subsubsection{Error propagation through Monte Carlo sampling}

\juan{We estimated the effect of the noise in the circular statistics introduced in Sec.~\ref{sec:circstats} by using Monte Carlo sampling.
We generated $10^{4}$ draws $I_{i}(l,b,v)$ from a Gaussian probability distribution function described by the mean value $I(l,b,v)$ and the variance $(\sigma_{I}$, where $I(l,b,v)$ are the observed intensities in the PPV cubes and $\sigma_{I}$ is estimated from \juan{noise-dominated} velocity channels. 
As a zeroth-order approximation, we assumed that the noise is constant in each velocity-channel map and across the velocity channels. 
This assumption is well justified by the uniform $uv$ coverage and total integration times throughout the THOR observations \citep{beuther2016,wang2020hi}.}

\juan{From each of the generated $I_{i}(l,b,v)$ we calculated $r$, $V$, and $\left<\theta\right>$.
Assuming that the errors in each of these quantities have a Gaussian distribution, we reported the mean value and the standard deviation for each on of them.
Given the selection by $I$ signal-to-noise introduced in Sec.~\ref{sec:filselection}, the estimated uncertainties obtained through this method are very small.
}

\subsubsection{Curvature threshold}

We calculated the curvature threshold values $\lambda^{C}_{-}$ introduced in Sec.~\ref{sec:filselection} by considering a velocity channel with very low H{\sc i} in each 2\deg\,$\times$\,2\deg\ region.
For that channel, we estimated the mean intensity and the maximum curvature $\lambda_{-}$, as defined in Eq.~\eqref{eq:lambda}, which we assign to be $\lambda^{C}_{-}$ for that particular position.
Figure~\ref{fig:NoiseChannel} shows an example of this procedure for the region presented in Fig.~\ref{fig:example}.
Given that the selected velocity channel is dominated by noise, the filamentary structures cover the whole maps and present values of $\lambda_{-}$ close to $0$, that is, very low curvature.
The orientation of these filaments is rather homogeneous and it is not indicative of the spatial distribution of the noise.

\subsubsection{Spatial structure of the noise}

To characterize the spatial structure of the noise, we use the noise map of the continuum emission at 1.42\,GHz, $\sigma_{I_{1.42}}$, presented in Fig.~\ref{fig:ContinuumNoise}.
The noise map of the continuum emission serves as a proxy for the linear structures that can be potentially introduced in the H{\sc i} maps by continuum sources in absorption.
By masking the H{\sc i} emission based on the filamentary structures found in $\sigma_{I_{1.42}}$, we exclude the strong continuum sources and the side-lobe features around them from the Hessian analysis.
We note that in general the orientation of the filamentary structure in $\sigma_{I_{1.42}}$ rarely corresponds to that found in the H{\sc i} emission, as we show in the example presented in \ref{fig:exampleCircStats}.
But that is not necessarily the case in all the 2\deg\,$\times$\,2\deg\ regions, which motivates our masking scheme.

\subsection{Derivative kernel size}

In the main body of this paper, we have shown the results for a particular selection of the derivative kernel size with 120\arcsec\ FWHM.
This selection, which sets the scale at which the filamentary structures are evaluated, was selected empirically by reaching a compromise between the spurious filamentary structures \juan{obtained} with a kernel size \juan{close to the resolution of the observations (40\arcsec)} and the loss of information resulting from using a very coarse one.

\begin{figure*}[ht!]
\centerline{\includegraphics[width=0.95\textwidth,angle=0,origin=c]{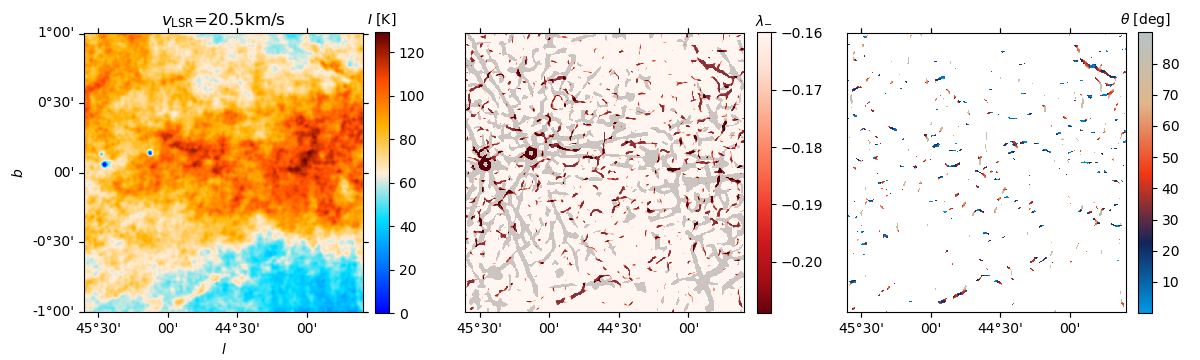}}
\vspace{-0.6cm}
\centerline{\includegraphics[width=0.95\textwidth,angle=0,origin=c]{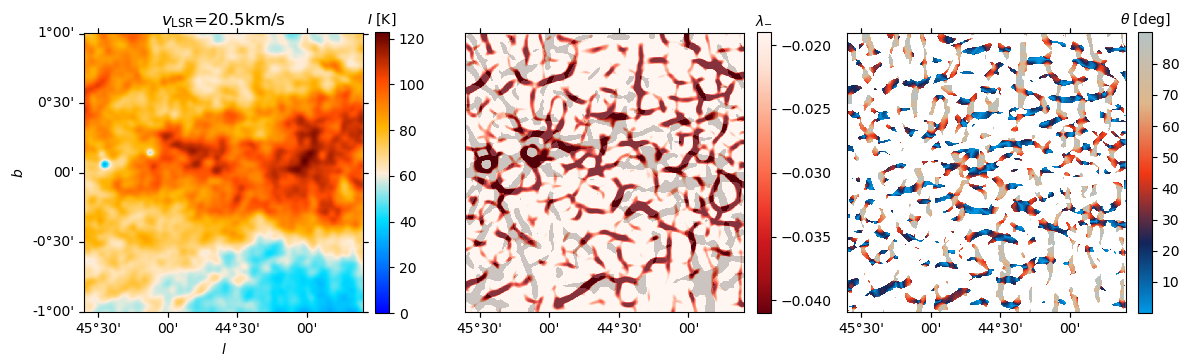}}
\caption{Same as Fig.~\ref{fig:example} for derivative kernels of 80\arcs\ (top) and 160\arcs\ FWHM (bottom)\juan{, which correspond to two and four times the native resolution of the \thorhi\ observations, respectively}.
}\label{fig:exampleMultiKernel}
\end{figure*}

Figure~\ref{fig:exampleMultiKernel} shows an example of two different kernels sizes applied to the same velocity channel map toward the region presented in Fig.~\ref{fig:example}.
\juan{The kernel that corresponds to twice the native resolution of the observations (80\arcsec\ FWHM)} highlights a large number of narrow filamentary structures, but it is very sensitive to the features of the H{\sc i} imaging.
Some of these features are the result of the artifacts from the interferometric data and are common when considering next-neighbour derivatives. 
The coarser 160\arcsec\ FWHM kernel, shows a much clearer contrast in terms of $\lambda_{-}$, but washes away some of the structures in the intensity map.
This selection may need further investigation for the study of other filament properties, such as the width, but it does not critically affect the results of the orientation study.

\begin{figure}[ht!]
\centerline{
\includegraphics[width=0.49\textwidth,angle=0,origin=c]{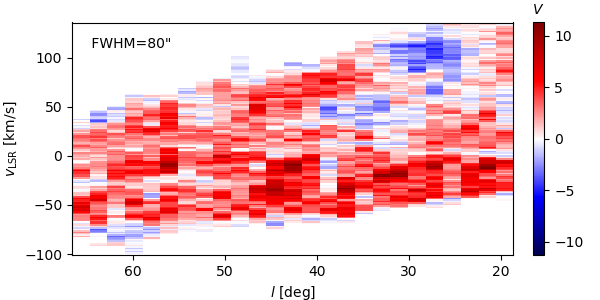}
}
\centerline{
\includegraphics[width=0.49\textwidth,angle=0,origin=c]{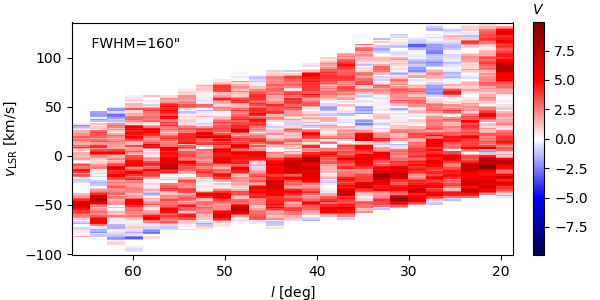}
}
\caption{Longitude-velocity diagram of the projected Rayleigh statistic $V$ estimated using the indicated derivative kernel sizes.
}\label{ig:lvdiagramsPRS}
\end{figure}

\begin{figure}[ht!]
\centerline{
\includegraphics[width=0.49\textwidth,angle=0,origin=c]{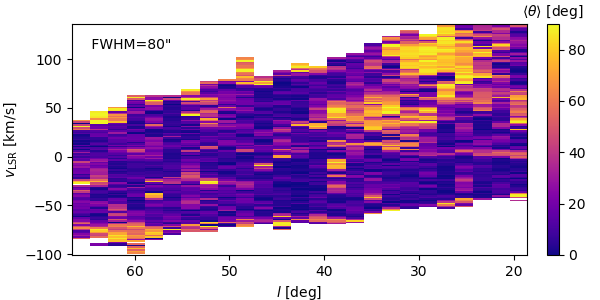}
}
\centerline{
\includegraphics[width=0.49\textwidth,angle=0,origin=c]{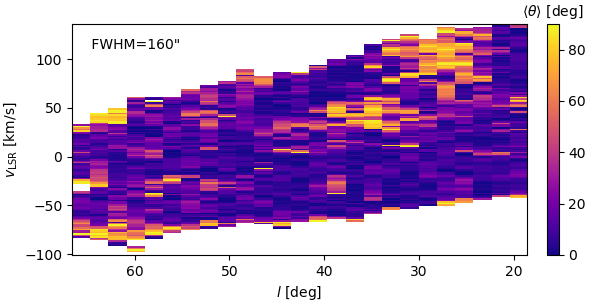}
}
\caption{Longitude-velocity diagram of the mean orientation angle $\left<\theta\right>$ of the H{\sc i} filamentary structures estimated using the Hessian method and the indicated derivative kernel sizes.
}\label{ig:lvdiagramsMeanTheta}
\end{figure}

The $lv$ diagrams of the $V$ and $\left<\theta\right>$ obtained with the 80\arcsec\ and 160\arcsec\ FWHM derivative kernels are shown in Fig.~\ref{ig:lvdiagramsPRS} and Fig.~\ref{ig:lvdiagramsMeanTheta}.
The results are in general similar to those presented in Fig.~\ref{fig:lvdiagrams}.
However, it is evident that the 80\arcs\ FWHM kernel appears noisier in both $V$ and $\theta$, most likely related to the effect of the spatial features shown in Fig.~\ref{fig:exampleMultiKernel}.
The 160\arcs\ FWHM kernel shows lower maximum levels of $V$, but the main regions of interest in Fig.~\ref{fig:PRSandHIIregions} are still clearly identifiable in the $lv$ diagrams.

\subsection{Filament selection}

\begin{figure}[ht!]
\centerline{\includegraphics[width=0.49\textwidth,angle=0,origin=c]{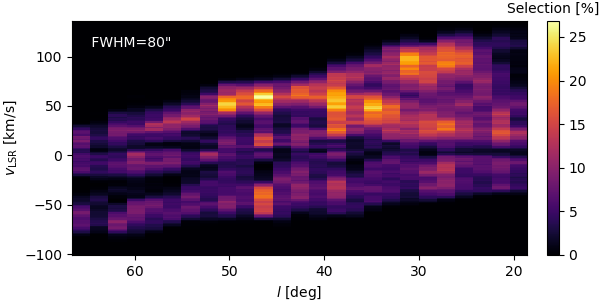}}
\centerline{\includegraphics[width=0.49\textwidth,angle=0,origin=c]{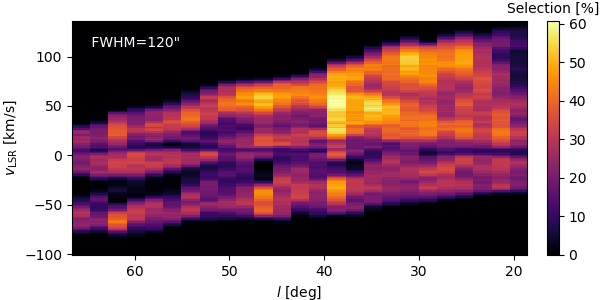}}
\centerline{\includegraphics[width=0.49\textwidth,angle=0,origin=c]{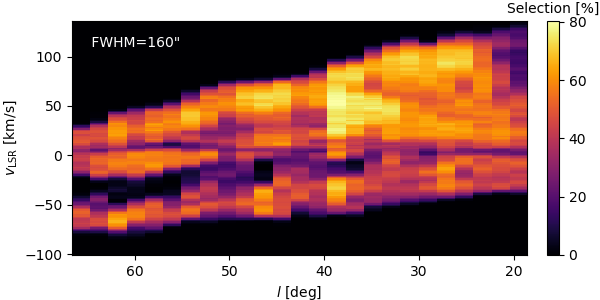}}
\caption{Longitude-velocity diagram showing the percentage of each velocity-channel map that is selected in the Hessian analysis.
Each pixel corresponds to a 2\deg\,$\times$\,2\deg\ tile. 
}\label{fig:lvdiagramsSelection}
\end{figure}

\juan{Besides the $I$ signal-to-noise considerations, the selection of the filaments is based on the curvature threshold $\lambda^{C}_{-}$, which is calculated independently for each derivative kernel size.
The area of each tile that is covered by the selected filaments changes with the derivative kernel selection.
For example, if we picked a very large derivative kernel, we would obtain a very large filament, which is the Galactic plane itself, that covers most of the tile.
As the kernel size is reduced one expects that the area of the tile covered by filaments is also reduced, as we are selecting smaller scales and sampling narrower filaments.
This expectation is confirmed for the \thorhi\ observations in Fig~\ref{fig:lvdiagramsSelection}.
} 

\juan{The fact that the percentage of the tile covered by filaments is in general changing with the size of the derivative kernel is an indication that most of them are not resolved at the spatial scales considered in Fig~\ref{fig:lvdiagramsSelection}.
If there was a particular scale at which the filaments are resolved, the map coverage would remain constant for derivative kernel sizes below that scale.
This is due to the fact that the curvature of a resolved structure, quantified by the value of $\lambda_{-}$, should be independent of the sampled scale.
This motivates further observations at higher resolution to study the width of the structures.
}

The \juan{percentage of a tile that is identified as filaments changes} across $l$ and $v_{\sc LSR}$ and roughly follows the same distribution of the mean intensity, shown in Fig.~\ref{fig:lvdiagrams}.
\juan{This suggests} that more filamentary structures are found in the highest H{\sc i} intensity tiles.
For the smallest derivative kernel, 80\arcsec\ FWHM, the selected filamentary structures correspond to up to 25\% of the area of the tiles but it can be up to 80\% in the case of the 160\arcsec\ FWHM kernel.
The selected percentage does not show any evident correlation with the orientation of the filamentary structures.

\section{Comparison with \galfahi}\label{app:GALFA}

\begin{figure*}[ht!]
\centerline{\includegraphics[width=0.99\textwidth,angle=0,origin=c]{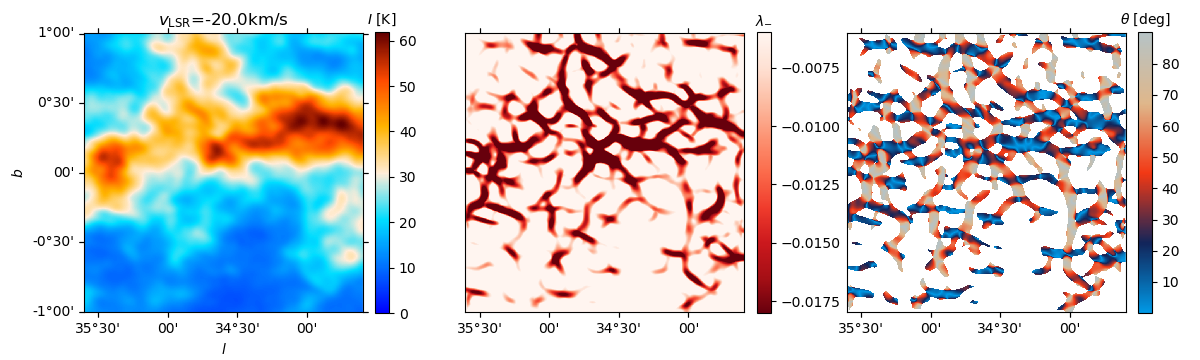}}
\centerline{\includegraphics[width=0.99\textwidth,angle=0,origin=c]{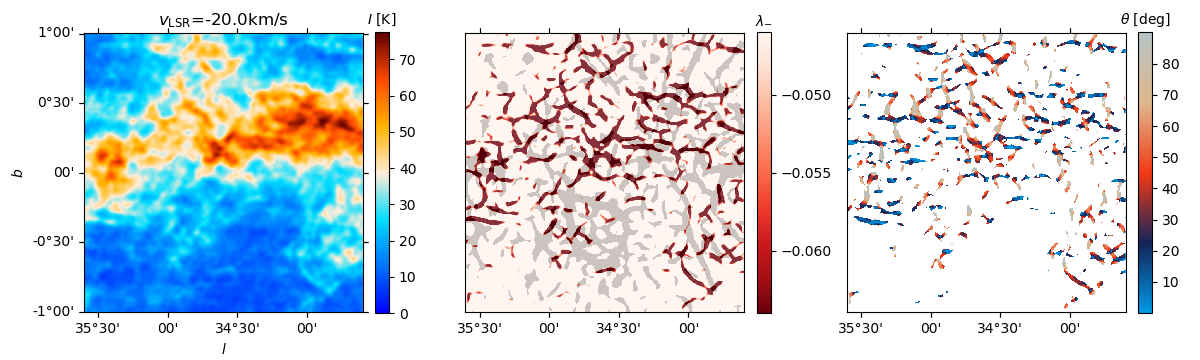}}
\caption{Examples of the results of the Hessian analysis applied to the GALFA H{\sc i} and THOR-H{\sc i} observations, shown in the top and bottom panels respectively.}
\label{fig:comparisonTHORandGALFA}
\end{figure*}

With an angular resolution of 4\arcmin, \galfahi\ is the highest resolution single-dish observation that we can use to evaluate potential artifacts introduced by the inteferometer and the validity of our masking scheme in the analysis of the \thorhi\ data.
We present an example of the Hessian analysis of a 2\deg\,$\times$\,2\deg\ and 1.5-\kps\ tile in both surveys in Fig.~\ref{fig:comparisonTHORandGALFA}.
\juan{Figure~\ref{fig:scatterVinTHORandGALFA} confirms that the Hessian analysis of the GALFA H{\sc i} and THOR-H{\sc i} observations gives very similar results when using the same derivative kernel size (240\arcmin) and the same criteria for filament selection}.

\begin{figure}[ht!]
\centerline{\includegraphics[width=0.49\textwidth,angle=0,origin=c]{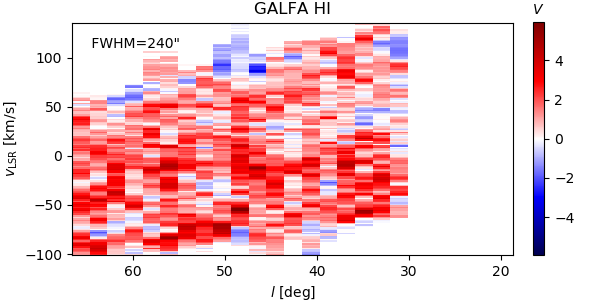}}
\vspace{0.2cm}
\centerline{\includegraphics[width=0.49\textwidth,angle=0,origin=c]{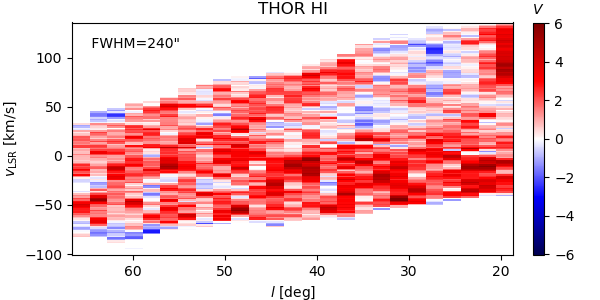}}
\caption{Longitude-velocity diagram of the projected Rayleigh statistic $V$ corresponding to the results of the Hessian analysis of the \galfahi\ and \thorhi\ observations, shown in the top and bottom.
Both results were obtained with the same derivative kernel size, indicated in the figure.}
\label{fig:multitracersPRSlvdiagram}
\end{figure}

\begin{figure}[ht!]
\centerline{
\includegraphics[width=0.49\textwidth,angle=0,origin=c]{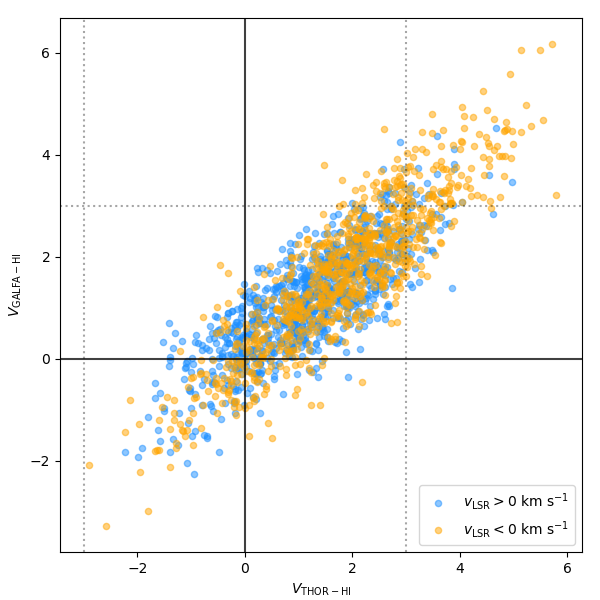}
}
\caption{\juan{Projected Rayleigh statistic $V$ calculated from the Hessian analysis of the \galfahi\ and \thorhi\ observations using a 240\arcsec\ FWHM derivative kernel.
Each point corresponds to a 2\deg\,$\times$\,2\deg\,$\times$\,1.5\,\kps\ velocity-channel tile fully covered by both surveys.
The vertical and horizontal dotted lines indicate $|V|$\,$=$\,3, which roughly corresponds to a 3-$\sigma$ confidence level in the estimation of a preferential orientation of 0\deg\ or 90\deg.}
}
\label{fig:scatterVinTHORandGALFA}
\end{figure}

Figure~\ref{fig:multitracersPRSlvdiagram} indicates that the vertical filaments identified in THOR-H{\sc i} around $l$\,$\approx$\,36\deg\ and $v_{\rm LSR}$\,$\approx$\,40\,km\,s$^{-1}$ are not prominent in GALFA-H{\sc i}.
It also indicates that the deviations from the general trend of $V$\,$>$\,0 around the maximum and minimum $v_{\rm LSR}$ are common to both datasets, but the higher sensitivity GALFA-H{\sc i} allows for their observation in more velocity channels.

\begin{figure}[ht!]
\centerline{\includegraphics[width=0.49\textwidth,angle=0,origin=c]{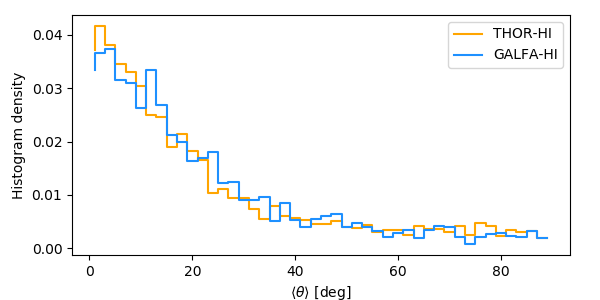}}
\caption{Histogram of the mean orientations of the H{\sc i} filaments identified using the Hessian analysis \juan{with a 240\arcsec\ FWHM derivative kernel applied to 2\deg\,$\times$\,2\deg\,$\times$\,1.5\,\kps\ velocity-channel tiles in} the \galfahi\ and \thorhi\ observations.
}
\label{fig:histoMeanTheta}
\end{figure}

\section{Other filament finding methods}\label{app:othermethods}

\begin{figure*}[ht!]
\centerline{\includegraphics[width=0.95\textwidth,angle=0,origin=c]{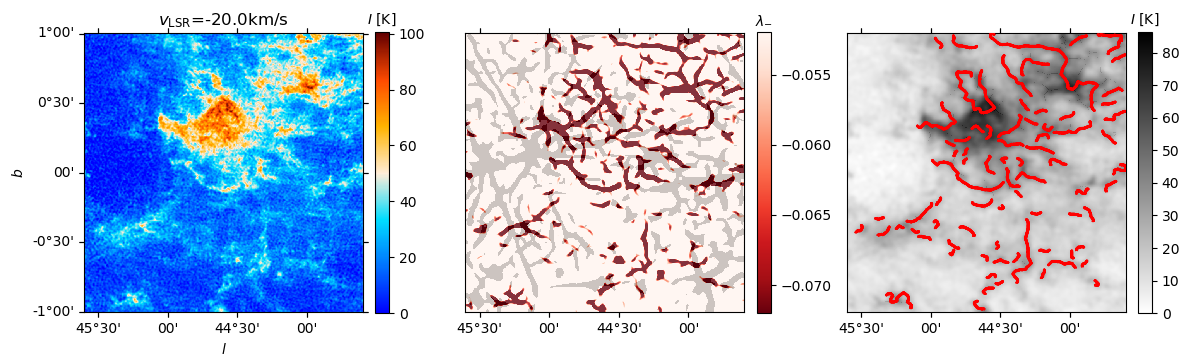}}
\centerline{\includegraphics[width=0.95\textwidth,angle=0,origin=c]{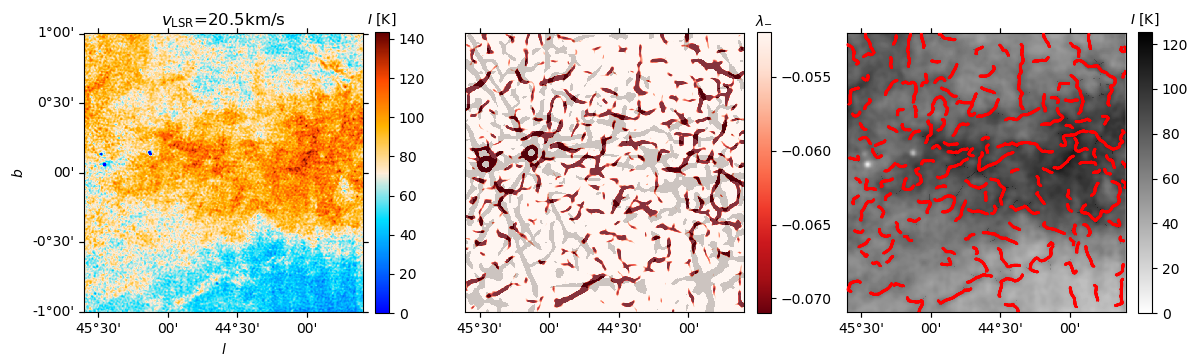}}
\caption{Example of the {\tt FilFinder} application to a pair of velocity channel maps in \thorhi.
{\it Left}. Intensity map at the indicated line-of-sight velocity.
{\it Middle}. Map of the eigenvalue $\lambda_{-}$ \juan{obtained with a 120\arcmin\ FWHM derivative kernel.}
\juan{This map is} used as a mask for {\tt FilFinder}.
{\it Right}. {\tt FilFinder} filaments overlaid on the respective H{\sc i} intensity map \juan{smoothed to 120\arcmin\ FWHM}.
}\label{fig:HESandFilFinderMaps}
\end{figure*}
\begin{figure*}[ht!]
\centerline{\includegraphics[width=0.95\textwidth,angle=0,origin=c]{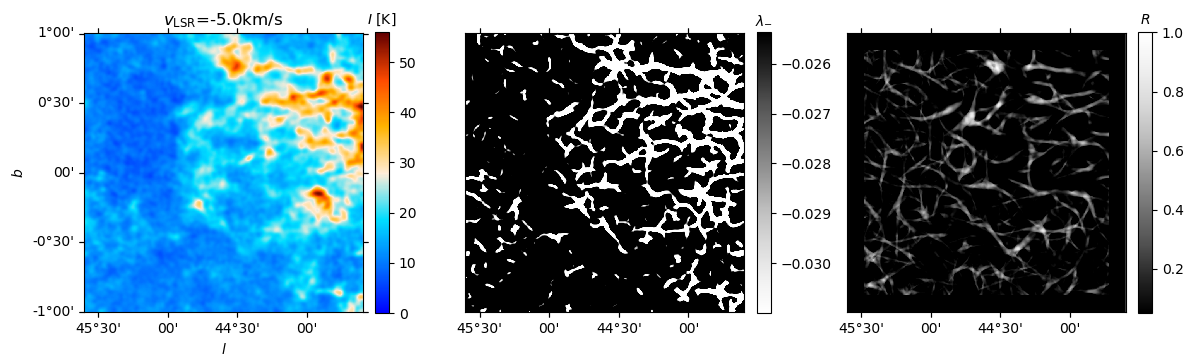}}
\centerline{\includegraphics[width=0.95\textwidth,angle=0,origin=c]{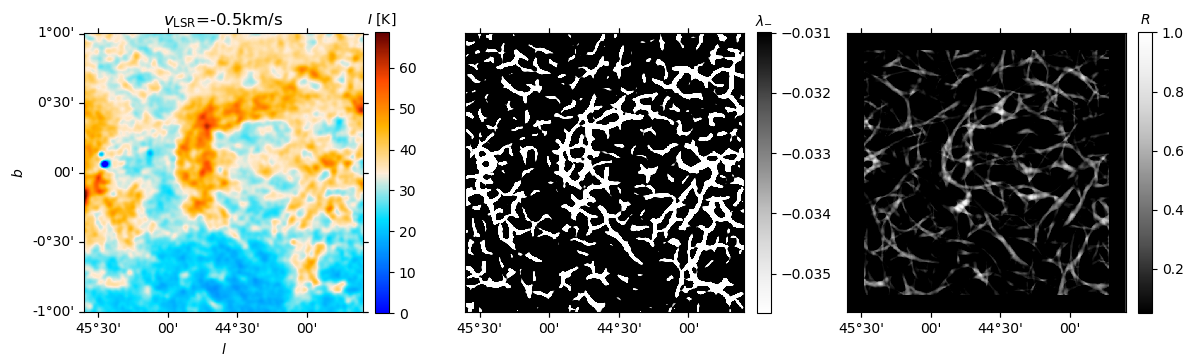}}
\caption{Comparison of the results of the Hessian and the RHT methods for two different velocity-channel maps toward a region within the area covered by \thorhi.
{\it Left}. Intensity map at the indicated line-of-sight velocity, $v_{\rm LSR}$.
{\it Middle}. Map of the eigenvalue $\lambda_{-}$ of the Hessian matrix.
{\it Right}. Map of the RHT backprojection.
}\label{fig:HESandRHTmaps}
\end{figure*}

Throughout this paper, we have used an algorithm for the identification of filamentary structures that is based on the Hessian matrix.
This is not the only filament finder algorithm, although it is considerably faster than other alternatives.
Additionally, the definition of its parameters (that is, the derivative kernel size and the curvature threshold) can be readily made in a self-consistent fashion.

In this section, we consider two alternative algorithms, {\tt FilFinder} and the Rolling Hough Transform, and show that their results are consistent with those found using the Hessian matrix.
These two algorithms are computationally more demanding, but they do not offer a significant advantage in the study of the orientation of the filamentary structures.
However, they are very powerful tools for studying properties such as the filament length and width that may be of interest in a follow-up study.

\subsection{{\tt FilFinder}}

{\tt FilFinder} is a {\tt Python} package for extraction and analysis of filamentary structure in molecular clouds introduced in \cite{koch2015}.
It segments filamentary structure by using adaptive thresholding. 
This thresholding is made over local neighborhoods, allowing for the extraction of structure over a large dynamic range. 
Using the filament mask, the length, width, orientation and curvature are calculated. 
Further features include extracting radial profiles along the longest skeleton path, creating filament-only model images, and extracting values along the skeleton.

However, one limiting restriction in the implementation of {\tt FilFinder} is the size of the map.
When applying {\tt FilFinder} to the 2\deg\,$\times$\,2\deg\ tiles in \thorhi, the memory requirements made it impractical to use it without careful masking of the maps.
We used a mask based on the curvature $\lambda_{-}$ obtained from the Hessian analysis to produce the example presented in Fig.~\ref{fig:HESandFilFinderMaps}.
Different masking schemes, such as those based on the intensity, will introduce bias toward filaments in channels with low intensity background.

Figure~\ref{fig:HESandFilFinderMaps} shows that {\tt FilFinder} highlights most of the elongated structures found with the Hessian algorithm.
It also finds significant connectivity among them, resulting in very long curved filaments. 
We conclude from this experiment with {\tt FilFinder} that the masking scheme would constitute the main source of discrepancy between this and other methods.
Given that we obtained the best results using a mask that is based on the curvature, {\tt FilFinder} does not produce a different result than that in the Hessian method.
Needless to say, {\tt FilFinder} constitutes a powerful tool to study other properties of the H{\sc i} filaments, such as their width or length, but those studies are beyond the scope of this work.

\subsection{The Rolling Hough transform method}

The Rolling Hough Transform \citep[RHT,][]{clark2014} is a tool for quantifying the linearity and spatial coherence of H{\sc i} structures.
In contrast to the Hessian matrix analysis, which is based on second order spatial derivatives, the RHT is a mapping between the image space $(x,y)$ and the space defined by the transformation
\begin{equation}
\rho = x\cos\theta + y\sin\theta,
\end{equation}
whose coordinates are $(\rho,\theta)$
Although the procedure described in \cite{clark2014} introduces additional steps aimed to select and evaluate image space features at a particular scale, at the core of the method is the 
transformation of each of the image points into a straight line in a parameter space. 
The output of the RHT is the function $R(\theta,x,y)$, which contains information on the directionality ($\theta$) of the image features in the position $(x,y)$.

In this section, we evaluate the difference in the results of the Hessian matrix and the RHT method by applying both techniques to the same area of the THOR H{\sc i} observations.
Figure~\ref{fig:HESandRHTmaps} shows the results of both analysis in a particular velocity channel.
The leftmost panel shows the intensity maps and the middle panel show the values of the curvature eigenvalues of the Hessian matrix ($\lambda_{-}$), introduced in \eqref{eq:lambda}.
The rightmost panel shows the RHT backprojection
\begin{equation}
R(x,y) = \int R(x,y,\theta),
\end{equation}
which is a visualization of the linear structures identified by the RHT.
At first glimpse, and without any selection based on the signal-to-noise ratio of the intensity map, both methods seem to trace the same structures. 

\begin{figure}[ht!]
\centerline{\includegraphics[width=0.49\textwidth,angle=0,origin=c]{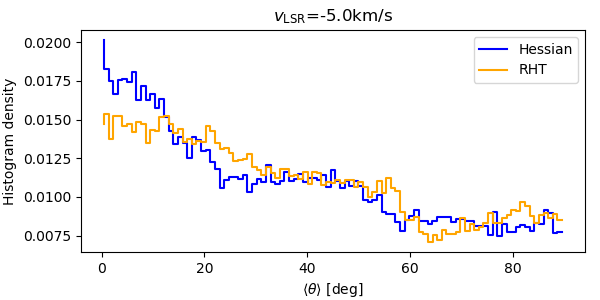}}
\centerline{\includegraphics[width=0.49\textwidth,angle=0,origin=c]{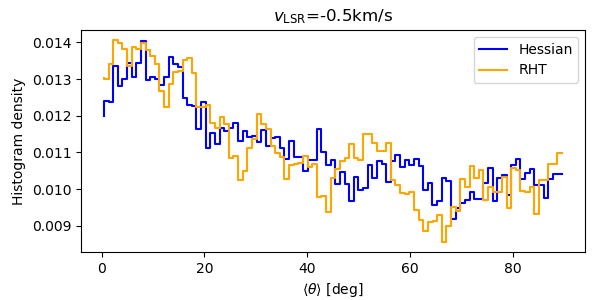}}
\caption{\juan{Histograms of the} orientation of the \juan{filamentary} structures identified using the Hessian matrix and the RHT methods {in the} velocity-channel maps presented in Fig.~\ref{fig:HESandRHTmaps}.
}\label{fig:HESandRHThist}
\end{figure}

\begin{figure}[ht!]
\centerline{
\includegraphics[width=0.49\textwidth,angle=0,origin=c]{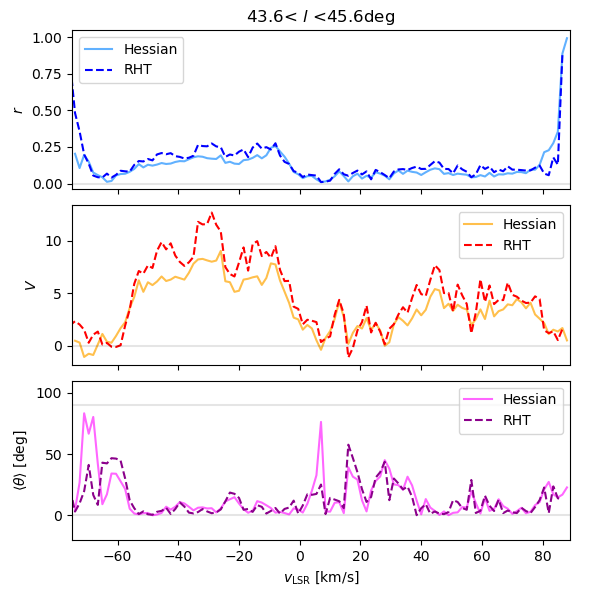}
}
\caption{Circular statistics used for the study of the orientation of the H{\sc i} filamentary structures found using the Hessian and the RHT methods across velocity channels toward the region presented in Fig.~\ref{fig:HESandRHTmaps}.
{\it Top}. Mean resulting vector ($r$), which indicates if the distribution of orientation angles is flat ($r$\,$\approx$\,0) or sharply unimodal ($r$\,$\approx$\,1).
{\it Middle}. Projected Rayleigh statistic ($V$), which indicates if the distribution of orientation angles is clearly peaked around 0\deg\ ($V$\,$\gg$\,\juan{0}) or 90\deg\ ($V$\,$\ll$\,\juan{0}).
{\it Bottom}. Mean orientation angle ($\left<\theta\right>$) of the H{\sc i} filamentary structures. 
}\label{fig:HESandRHTacrossVLSR}
\end{figure}

Figure~\ref{fig:HESandRHThist} shows the H{\sc i} filament orientation obtained using the Hessian method and the RHT in the tiles presented in Fig.~\ref{fig:HESandRHTmaps}.
The great similarity in the histograms of the orientation angles obtained with the two methods confirms that there is no significant difference between the global results obtained with either of them.
This is further confirmed in Fig.~\ref{fig:HESandRHTacrossVLSR}, which shows the results for both methods across velocity channels toward the same region.

The description of the filamentary structure in the Hessian and RHT methods is different.
While the Hessian matrix offers a characterization of the topology of the 2D scalar field, intensity map, the RHT describes the filamentary structures in that scalar field as straight line segments, as shown in \ref{fig:HESandRHTmaps}.
However, this fundamental difference does not produce a significant difference in the distribution of orientations of the filamentary structures.

\section{FRIGG simulations}\label{app:FRIGG}

\begin{figure*}[ht!]
\centerline{
\includegraphics[width=0.44\textwidth,angle=0,origin=c]{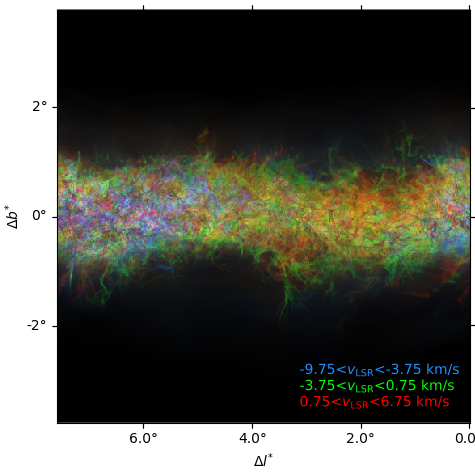}
\includegraphics[width=0.44\textwidth,angle=0,origin=c]{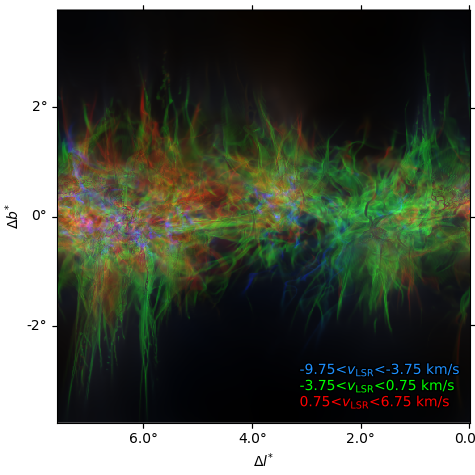}
}
\caption{
Synthetic observations of the H{\sc i} emission from the FRIGG simulations with a standard (3\,$\mu$G) and very high (12\,$\mu$G) initial magnetizations, shown in the left and right panels, respectively.
These snapshots are taken 14\,Myr before those shown in Fig.~\ref{fig:FRIGGrgb}.
The colors represent the emission in the three indicated radial velocity bins with the same average emission.
}\label{fig:FRIGGrgbAPP}
\end{figure*}
\begin{figure*}[ht!]
\centerline{
\includegraphics[width=0.49\textwidth,angle=0,origin=c]{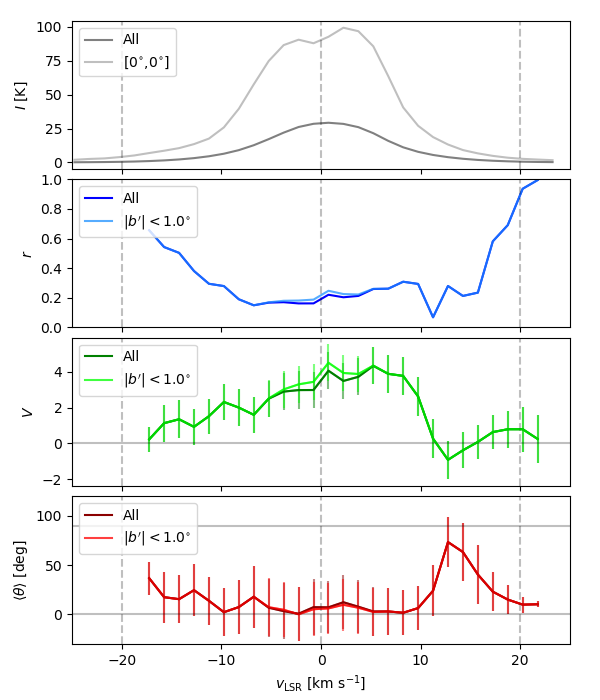}
\includegraphics[width=0.49\textwidth,angle=0,origin=c]{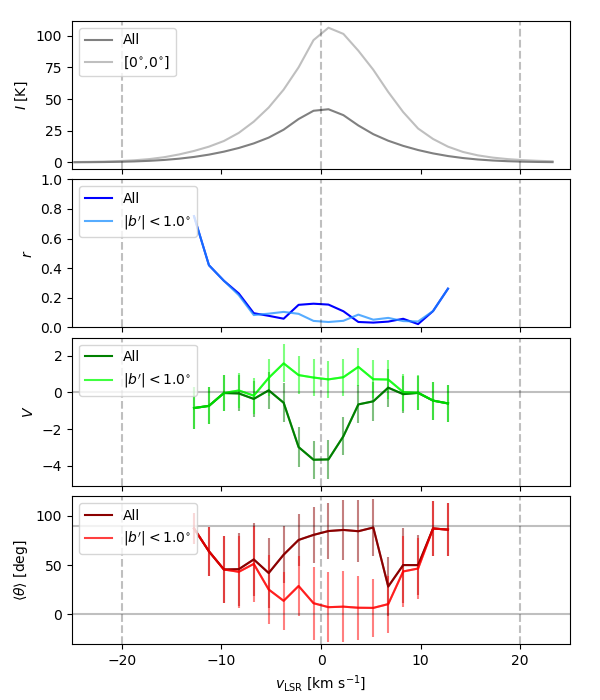}
}
\caption{\juan{Synthetic H{\sc i} emission intensity and} circular statistics used for the study of the orientation of the H{\sc i} filamentary structures across velocity channels in the synthetic observations presented in Fig.~\ref{fig:FRIGGrgbAPP}.
{\it Top}. \juan{H{\sc i} emission averaged over the whole map and toward the indicated position.}
{\it Middle top}. Mean resulting vector ($r$), which indicates if the distribution of orientation angles is flat ($r$\,$\approx$\,0) or sharply unimodal ($r$\,$\approx$\,1).
{\it Middle bottom}. Projected Rayleigh statistic ($V$), which indicates if the distribution of orientation angles is clearly peaked around 0\deg\ ($V$\,$\gg$\,\juan{0}) or 90\deg\ ($V$\,$\ll$\,\juan{0}).
{\it Bottom}. Mean orientation angle ($\left<\theta\right>$) of the H{\sc i} filamentary structures.
}\label{fig:FRIGGresultsAPP}
\end{figure*}


We produced synthetic H{\sc i} emission observations following the procedure described in appendix~C of \cite{soler2019} for a line of sight (LOS) that is perpendicular to the initial direction of the magnetic field.
\juan{We followed the general radiative transfer equations presented in \cite{draine2011}. 
We calculated the emissivity 
\begin{equation}
j(v_{\rm LSR}) = \frac{3}{16\pi}h\nu_{0}A_{10}\,n_{\rm H}\,\phi(v_{\rm LSR})
\end{equation}
and absorption coefficient
\begin{equation}\label{eq:kappa}
\kappa(v_{\rm LSR}) = \frac{3}{32\pi}\frac{hc^{2}}{k_{\rm B}\nu_{0}}A_{10}\frac{n_{\rm H}}{T_{\rm s}}\phi(v_{\rm LSR}),
\end{equation}
where $\nu_{0}$\,$=$\,1.4204\,GHz is the frequency at the line center, $A_{10}$\,$=$\,2.876\,$\times$\,10$^{-15}$\,s$^{-1}$ is the Einstein coefficient of the line, $n_{\rm H}$ is the H{\sc i} number density, and $\phi(v_{\rm LSR})$ is the normalized line profile as a function of velocity.
We assumed a Gaussian velocity distribution
\begin{equation}
\phi(v_{\rm LSR}) = \frac{1}{\sqrt{\pi}\Delta v}\frac{c}{\nu_{0}}e^{-[(v_{\rm LSR}-v)/\Delta v]^{2}},
\end{equation}
where $\Delta v$\,$\equiv$\,$(2k_{\rm B}T_{\rm k}/m_{\rm H})^{1/2}$ is the Doppler width of the line.
Given that the FRIGG simulations do not include a chemical network, we assumed $n_{H}$ to be 60\,\% of the gas density in each cell.
This rough assumption, which is based on the average values in the Milky Way at galactocentric distances $R$\,$<$\,20\,kpc \citep{draine2011}, does not have a critical effect in the results of our study of this numerical experiment.}

\juan{We assumed that the excitation temperature (spin temperature, $T_{\rm s}$) of the 21-cm line is the same as the kinetic temperature of the gas, $T_{\rm k}$.
This assumption leads to an over estimation of $T_{\rm s}$ in warm gas \citep[see for example][and the references therein]{kim2014}, but it is sufficient for the current exploration of the simulations.
We used parallel ray-tracing assuming a far-field geometry.
We calculated the absorption coefficient using Eq.~\eqref{eq:kappa} and the contribution to the local optical depth from the $i$-th simulation element along the LOS, which is $\tau^{i}(v_{\rm LSR})$\,$=$\,$\kappa(v_{\rm LSR}; s_{i})\Delta s$, where $s_{i}$\,$=$\,$i\Delta s$ and $\Delta s$\,$=$\,1.95\,pc.
The total optical depth along an LOS is
\begin{equation}
\tau(v_{\rm LSR})=\sum^{N}_{i=1}\tau^{i}(v_{\rm LSR}),
\end{equation}
where $N$ is the total number of LOS element in a given $l,b$ direction.
The synthetic H{\sc i} emission is the result of the sum of the brightness temperature of each element and the foreground attenuation, such that
\begin{equation}
T_{B}(v_{\rm LSR})=\sum^{N}_{i=1}\left[T_{\rm s}^{i}\left(1-e^{-\tau^{i}(v_{\rm LSR})}\right)\exp\left(-\sum^{i-1}_{j=1}\tau^{i}(v_{\rm LSR})\right)\right].
\end{equation}
The FRIGG simulations 

We calculated the emission between $-30$ and $30$\,\kps\ in velocity channels of 0.5\,\kps\ width, which we smoothed and regridded to match the velocity resolution of the \thorhi\ observations.
The output PPV cubes contain 512$^2$ pixels in the spatial dimensions, which we convolve with a two-dimensional Gaussian function with 40\arcs\ FWHM to match the angular resolution of the \thorhi\ observations.}

The front domain of the simulations is located 7.5\,kpc away to match the kinematic distance of the features found at $l$\,$\approx$\,20\deg\ and $v_{\rm LSR}$\,$=$\,120\,km\,s$^{-1}$.
Given that these simulations do not include a chemical network, we use thresholds in density and temperature to define the regions that produce H{\sc i} emission.
Explicitly, we consider the emission from the simulation voxels with particle densities $n$\,$<$\,50\,cm$^{-3}$ and temperatures in the range 20\,$<$\,$T$\,$<$\,$3$\,$\times$\,10$^{4}$\,K, which correspond to the general range of physical conditions where atomic hydrogen is expected \citep[see for example,][]{walch2015}.
Although these thresholds provide a very rough approximation to the real conditions in the ISM, they are sufficient for the numerical experiment that we are proposing here.

The synthetic H{\sc i} observations, shown in Fig.~\ref{fig:FRIGGrgb}, reveal a plethora of filamentary structures in the emission from both simulations.
In the simulation with standard magnetization, these filaments form an intricate network that extends up to roughly 125\,pc above and below the midplane.
In the simulation with very high magnetization, these filaments are much longer, they appear perpendicular to the plane both below and above it, and extend further that in the standard magnetization case.
In both cases, these filaments mostly correspond to density structures, as shown in the integrated density plots in \cite{iffrig2017}, rather that to fluctuations in the velocity field.
For the sake of comparison with the results presented in Sec.~\ref{sec:mhdsimsFRIGG}, we present in Fig.~\ref{fig:FRIGGrgbAPP} and Fig.~\ref{fig:FRIGGresultsAPP} the results of the synthetic observations and Hessian analysis of the standard and high-magnetization simulations in snapshots taken at 50 and 66\,Myr, respectively.

\section{Cloud Factory simulations}\label{app:CloudFactory}

We used the cubes of atomic hydrogen resulting from the time-dependent chemical model in the {\tt CloudFactory} simulations to perform synthetic observations using the {\tt lime} radiative transfer code \citep{brinch2010}.
The modifications involve the optimisation of the code and its adaptation to deal with simulated unstructured meshes in the second paper of the {\tt CloudFactory} series \citep{izquierdo2020submitted}.
Prior to running {\tt lime}, we use the {\tt grid} and {\tt rt} tools of the {\tt sf3dmodels} package to read, process and provide format to the simulation snapshots \citep{izquierdo2018}. 

The H{\sc i} 21\,cm hyperfine line has a low emission coefficient $A_{10}$\,$=$\,2.876\,$\times$\,10$^{-15}$\,s$^{-1}$ and hence it is easily populated by collisions even in the diffuse interstellar medium of the Galaxy \citep{purcellANDfield1956}. 
For this reason, it is sufficient to assume an LTE approximation to solve the radiation transport in this scenario, though {\tt lime} provides as well an iterative non-LTE implementation suitable for different gas tracers and physical conditions. 
For the level populations, the statistical weights of the upper and lower spin states are $g_{1}$\,$=$\,3 (triplet) and $g_{0}$\,$=$\,1 (singlet state), respectively. 
We are assuming that the spin temperature ($T_{\rm s}$) of the H{\sc i} tracks the gas kinetic temperature ($T_{\rm k}$), which is an acceptable approximation for $T_{\rm k}$\,$<$\,5\,$\times$\,$10^{3}$\,K \citep[see for example,][]{kim2014}.
For $T_{\rm k}$\,$>$\,5\,$\times$\,$10^{3}$\,K, this assumption leads to an over estimation of the spin temperature and the linewidths of the H{\sc i} emission that does not significantly affect the results of our study.

\juan{We calculated the emission between $-50$ and $50$\,\kps\ in velocity channels of 0.5\,\kps\ width, which we smoothed and regridded to match the velocity resolution of the \thorhi\ observations}.
The output PPV cubes contain 512$^2$ pixels \juan{in the spatial dimensions, which we convolve with a two-dimensional Gaussian function with 40\arcs\ FWHM to match the angular resolution of the \thorhi\ observations}.
\juan{As in the FRIGG simulations, the} front domain of the simulations is located 7.5\,kpc away.

\begin{figure*}[ht!]
\centerline{
\includegraphics[width=0.44\textwidth,angle=0,origin=c]{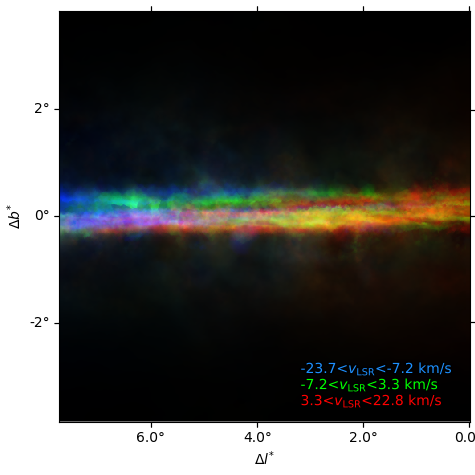}
\includegraphics[width=0.44\textwidth,angle=0,origin=c]{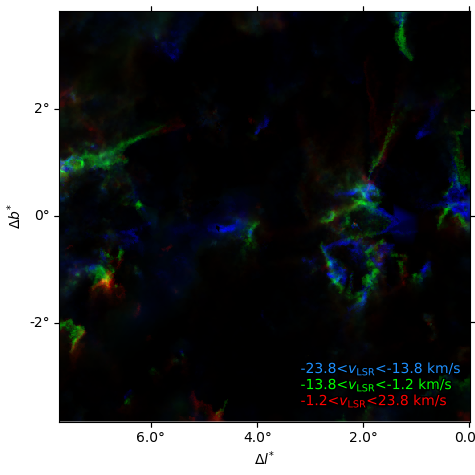}
}
\caption{
Synthetic observations of the H{\sc i} emission from the potential and feedback dominated {\tt CloudFactory} simulations, presented in the left and right panels, respectively.
These results correspond to a line of sight at 90\deg\ from that considered in the synthetic observations presented in Fig~\ref{fig:CloudFactoryRGB}.
The colors represent the emission in the three indicated radial velocity bins, each of them with the same average emission.
}\label{fig:CloudFactoryRGB90}
\end{figure*}

\begin{figure*}[ht!]
\centerline{
\includegraphics[width=0.49\textwidth,angle=0,origin=c]{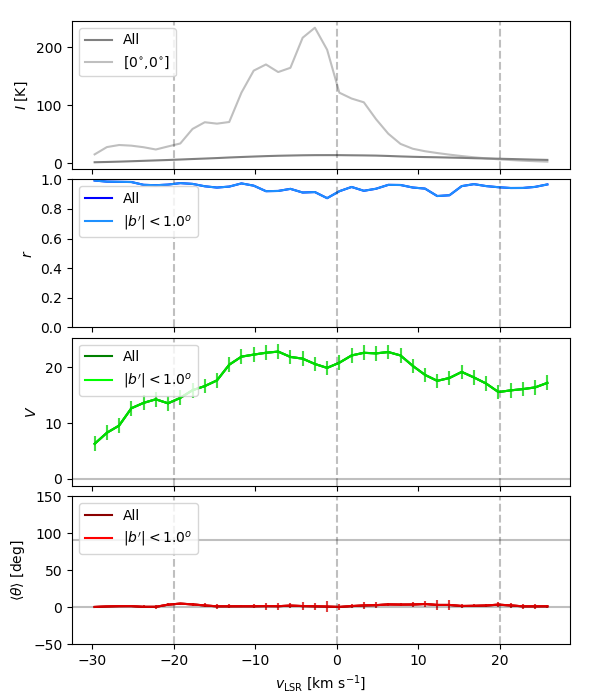}
\includegraphics[width=0.49\textwidth,angle=0,origin=c]{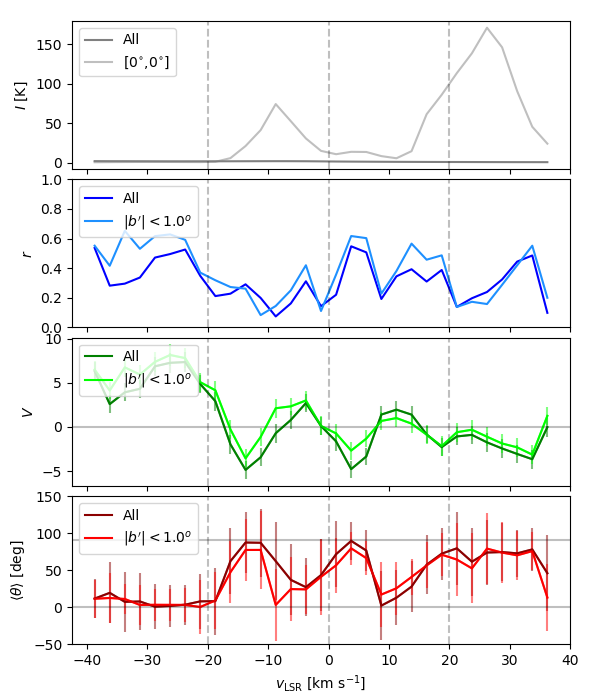}
}
\caption{\juan{Synthetic H{\sc i} emission intensity and} circular statistics used for the study of the orientation of the H{\sc i} filamentary structures across velocity channels in the synthetic observations presented in Fig.~\ref{fig:CloudFactoryRGB90}.
{\it Top}. \juan{H{\sc i} emission averaged over the whole map and toward the indicated position.}
{\it Middle top}. Mean resulting vector ($r$), which indicates if the distribution of orientation angles is flat ($r$\,$\approx$\,0) or sharply unimodal ($r$\,$\approx$\,1).
{\it Middle bottom}. Projected Rayleigh statistic ($V$), which indicates if the distribution of orientation angles is clearly peaked around 0\deg\ ($V$\,$\gg$\,\juan{0}) or 90\deg\ ($V$\,$\ll$\,\juan{0}).
{\it Bottom}. Mean orientation angle ($\left<\theta\right>$) of the H{\sc i} filamentary structures.
}\label{fig:CloudFactoryResults90}
\end{figure*}

We produced two edge-on views of the simulation snapshots to explore variations in the relative orientation of H{\sc i} filamentary structures.
The line of sight orthogonal to the Galactic rotation is presented in Fig.~\ref{fig:CloudFactoryRGB}.
The line of sight tangent to the Galactic rotation is presented in Fig.~\ref{fig:CloudFactoryRGB90}.
The line of sight selection does not produce a significant difference in the orientation of the H{\sc i} filaments, as illustrated in Fig.\ref{fig:CloudFactoryResults90}.


\end{document}

%% file: Planck.tex
\def\setsymbol#1#2{\expandafter\def\csname #1\endcsname{#2}}
\def\getsymbol#1{\csname #1\endcsname}

\def\all2013resultspapers{\nocite{planck2013-p01, planck2013-p02, planck2013-p02a, planck2013-p02d, planck2013-p02b, planck2013-p03, planck2013-p03c, planck2013-p03f, planck2013-p03d, planck2013-p03e, planck2013-p01a, planck2013-p06, planck2013-p03a, planck2013-pip88, planck2013-p08, planck2013-p11, planck2013-p12, planck2013-p13, planck2013-p14, planck2013-p15, planck2013-p05b, planck2013-p17, planck2013-p09, planck2013-p09a, planck2013-p20, planck2013-p19, planck2013-pipaberration, planck2013-p05, planck2013-p05a, planck2013-pip56, planck2013-p06b, planck2013-p01a}}

\newbox\tablebox    \newdimen\tablewidth
\def\leaderfil{\leaders\hbox to 5pt{\hss.\hss}\hfil}
\def\tablenote#1 #2\par{\begingroup \parindent=0.8em
    \abovedisplayshortskip=0pt\belowdisplayshortskip=0pt
    \noindent
    $$\hss\vbox{\hsize\tablewidth \hangindent=\parindent \hangafter=1 \noindent
    \hbox to \parindent{$^#1$\hss}\strut#2\strut\par}\hss$$
    \endgroup}

\def\L2{\ifmmode L_2\else $L_2$\fi}

\def\DeltaT{\ifmmode \Delta T\else $\Delta T$\fi}
\def\deltat{\ifmmode \Delta t\else $\Delta t$\fi}
\def\fknee{\ifmmode f_{\rm knee}\else $f_{\rm knee}$\fi}
\def\Fmax{\ifmmode F_{\rm max}\else $F_{\rm max}$\fi}
\def\solar{\ifmmode{\rm M}_{\mathord\odot}\else${\rm M}_{\mathord\odot}$\fi}
\def\Msolar{\ifmmode{\rm M}_{\mathord\odot}\else${\rm M}_{\mathord\odot}$\fi}
\def\Lsolar{\ifmmode{\rm L}_{\mathord\odot}\else${\rm L}_{\mathord\odot}$\fi}
\def\inv{\ifmmode^{-1}\else$^{-1}$\fi}
\def\mo{\ifmmode^{-1}\else$^{-1}$\fi}
\def\sup#1{\ifmmode ^{\rm #1}\else $^{\rm #1}$\fi}
\def\expo#1{\ifmmode \times 10^{#1}\else $\times 10^{#1}$\fi}
\def\,{\thinspace}
\def\lsim{\mathrel{\raise .4ex\hbox{\rlap{$<$}\lower 1.2ex\hbox{$\sim$}}}}
\def\gsim{\mathrel{\raise .4ex\hbox{\rlap{$>$}\lower 1.2ex\hbox{$\sim$}}}}

\def\simprop{\mathrel{\raise .4ex\hbox{\rlap{$\propto$}\lower 1.2ex\hbox{$\sim$}}}}
\def\deg{\ifmmode^\circ\else$^\circ$\fi}
\def\pdeg{\ifmmode $\setbox0=\hbox{$^{\circ}$}\rlap{\hskip.11\wd0 .}$^{\circ}
          \else \setbox0=\hbox{$^{\circ}$}\rlap{\hskip.11\wd0 .}$^{\circ}$\fi}
\def\arcs{\ifmmode {^{\scriptstyle\prime\prime}}
          \else $^{\scriptstyle\prime\prime}$\fi}
\def\arcm{\ifmmode {^{\scriptstyle\prime}}
          \else $^{\scriptstyle\prime}$\fi}
\newdimen\sa  \newdimen\sb
\def\parcs{\sa=.07em \sb=.03em
     \ifmmode \hbox{\rlap{.}}^{\scriptstyle\prime\kern -\sb\prime}\hbox{\kern -\sa}
     \else \rlap{.}$^{\scriptstyle\prime\kern -\sb\prime}$\kern -\sa\fi}
\def\parcm{\sa=.08em \sb=.03em
     \ifmmode \hbox{\rlap{.}\kern\sa}^{\scriptstyle\prime}\hbox{\kern-\sb}
     \else \rlap{.}\kern\sa$^{\scriptstyle\prime}$\kern-\sb\fi}
\def\ra[#1 #2 #3.#4]{#1\sup{h}#2\sup{m}#3\sup{s}\llap.#4}
\def\dec[#1 #2 #3.#4]{#1\deg#2\arcm#3\arcs\llap.#4}
\def\deco[#1 #2 #3]{#1\deg#2\arcm#3\arcs}
\def\rra[#1 #2]{#1\sup{h}#2\sup{m}}

\def\dots{\relax\ifmmode \ldots\else $\ldots$\fi}
\def\WHzsr{\ifmmode $W\,Hz\mo\,sr\mo$\else W\,Hz\mo\,sr\mo\fi}
\def\mHz{\ifmmode $\,mHz$\else \,mHz\fi}
\def\GHz{\ifmmode $\,GHz$\else \,GHz\fi}
\def\mKs{\ifmmode $\,mK\,s$^{1/2}\else \,mK\,s$^{1/2}$\fi}
\def\muKs{\ifmmode \,\mu$K\,s$^{1/2}\else \,$\mu$K\,s$^{1/2}$\fi}
\def\muKRJs{\ifmmode \,\mu$K$_{\rm RJ}$\,s$^{1/2}\else \,$\mu$K$_{\rm RJ}$\,s$^{1/2}$\fi}
\def\muKHz{\ifmmode \,\mu$K\,Hz$^{-1/2}\else \,$\mu$K\,Hz$^{-1/2}$\fi}
\def\MJysr{\ifmmode \,$MJy\,sr\mo$\else \,MJy\,sr\mo\fi}
\def\MJysrmK{\ifmmode \,$MJy\,sr\mo$\,mK$_{\rm CMB}\mo\else \,MJy\,sr\mo\,mK$_{\rm CMB}\mo$\fi}
\def\microns{\ifmmode \,\mu$m$\else \,$\mu$m\fi}

\def\muK{\ifmmode \,\mu$K$\else \,$\mu$\hbox{K}\fi}
\def\microK{\ifmmode \,\mu$K$\else \,$\mu$\hbox{K}\fi}
\def\muW{\ifmmode \,\mu$W$\else \,$\mu$\hbox{W}\fi}
\def\kms{\ifmmode $\,km\,s$^{-1}\else \,km\,s$^{-1}$\fi}
\def\kmsMpc{\ifmmode $\,\kms\,Mpc\mo$\else \,\kms\,Mpc\mo\fi}
\providecommand{\sorthelp}[1]{}